\documentclass[reprint,aps,prb,twocolumn,groupedaddress]{revtex4-2}
\usepackage{graphicx}
\usepackage{here}
\usepackage{amsmath}
\usepackage{braket}
\usepackage{amsfonts}
\usepackage{subcaption}
\usepackage{blkarray}
\usepackage{makecell}
\usepackage{ragged2e}
\usepackage{bm}

\usepackage{xcolor}

\begin{document}

\title{Spinless electric toroidal multipoles in ferroaxial ${\rm K_2Zr(PO_4)_2}$ revealed by symmetry-adapted closest Wannier analysis}
\author{Yu Xie, Rikuto Oiwa, and Satoru Hayami}
\affiliation{Graduate School of Science, Hokkaido University, Sapporo 060-0810, Japan}
\date{\today}

\begin{abstract}
From a symmetry perspective, ferroaxial order belongs to the same symmetry as time-reversal-even pseudovectors. 
Experimentally, ${\rm K_2Zr(PO_4)_2}$ is known to undergo a displacive-type phase transition from a non-ferroaxial to a ferroaxial phase. 
To identify the key microscopic ingredients driving this transition, we carry out a quantitative analysis combining density-functional theory calculations and symmetry-adapted closest Wannier analysis.
As a result, we show that electric toroidal dipole, electric toroidal octupole, and electric hexadecapole, which belong to the same irreducible representation, make dominant contributions to the ferroaxial transition. 
In particular, we find that spinless electric toroidal octupoles, which originate from spin-independent off-diagonal real hopping between the $p$ orbitals on P and O atoms and between the $d$ orbitals on Zr atoms and $p$ orbitals on O atoms, provide the most significant contributions. 
Moreover, we explicitly analyze the orbital characters involved in the relevant hybridizations associated with these multipoles. 
We further show that the relativistic spin--orbit coupling has a negligible influence on the ferroaxial transition.
These results demonstrate that spin-independent orbital hybridization between different orbitals on different atoms plays a crucial role in inducing the ferroaxial transition.
\end{abstract}
\maketitle

\section{INTRODUCTION}\label{sec_intro}
Ferroaxiality arises from the breaking of certain mirror symmetries while preserving both time-reversal ($\mathcal{T}$) and spatial-inversion ($\mathcal{P}$) symmetries~\cite{jin2020observation}. 
Such space-time inversion properties stand in sharp contrast to those of ferromagnetism, which breaks $\mathcal{T}$ symmetry, and ferroelectricity, which breaks $\mathcal{P}$ symmetry~\cite{Hlinka_PhysRevLett.113.165502, Hlinka_PhysRevLett.116.177602}.
In this sense, ferroaxiality is neither magnetic nor electric in nature, and at first glance, it may appear to be a rather featureless state. 

The experimental identification of ferroaxial transition in RbFe(MoO$_4$)$_2$ has opened up an avenue of research~\cite{jin2020observation, Owen_PhysRevB.103.054104, Guo_PhysRevB.107.L180102, zeng2025photo}. 
Indeed, ferroaxial order has since been discovered in the materials NiTiO$_3$~\cite{hayashida2020visualization, Hayashida_PhysRevMaterials.5.124409, yokota2022three, fang2023ferrorotational}, Ca$_5$Ir$_3$O$_{12}$~\cite{Hasegawa_doi:10.7566/JPSJ.89.054602, hanate2021first, hanate2023space, hayami2023cluster}, BaCoSiO$_4$~\cite{Xu_PhysRevB.105.184407}, Na$_2$Hf(BO$_3$)$_2$~\cite{nagai2023chemicalSwitching}, Na-superionic conductors~\cite{nagai2023chemical}, MnTiO$_3$~\cite{Sekine_PhysRevMaterials.8.064406}, and EuTe$_3$~\cite{singh2025ferroaxial}. 
In parallel, theoretical studies have uncovered two characteristic classes of physical phenomena associated with ferroaxial ordering: The first involves off-diagonal responses of conjugated physical quantities, including antisymmetric thermopolarization~\cite{Nasu_PhysRevB.105.245125}, longitudinal spin current generation~\cite{Roy_PhysRevMaterials.6.045004, Hayami_doi:10.7566/JPSJ.91.113702}, and nonlinear transverse magnetization~\cite{inda2023nonlinear, du2025electrotoroidicity};
The second concerns cross correlations between chiral and polar degrees of freedom~\cite{cheong2021permutable, cheong2022linking, hayami2023chiral}, exemplified by the electrogyration effect~\cite{hayashida2020visualization, Martinez2025ferroaxial} and surface-induced chiral-domain selection~\cite{hayami2025chirality}.

\begin{figure}[htbp]
 \includegraphics[width=0.9\linewidth]{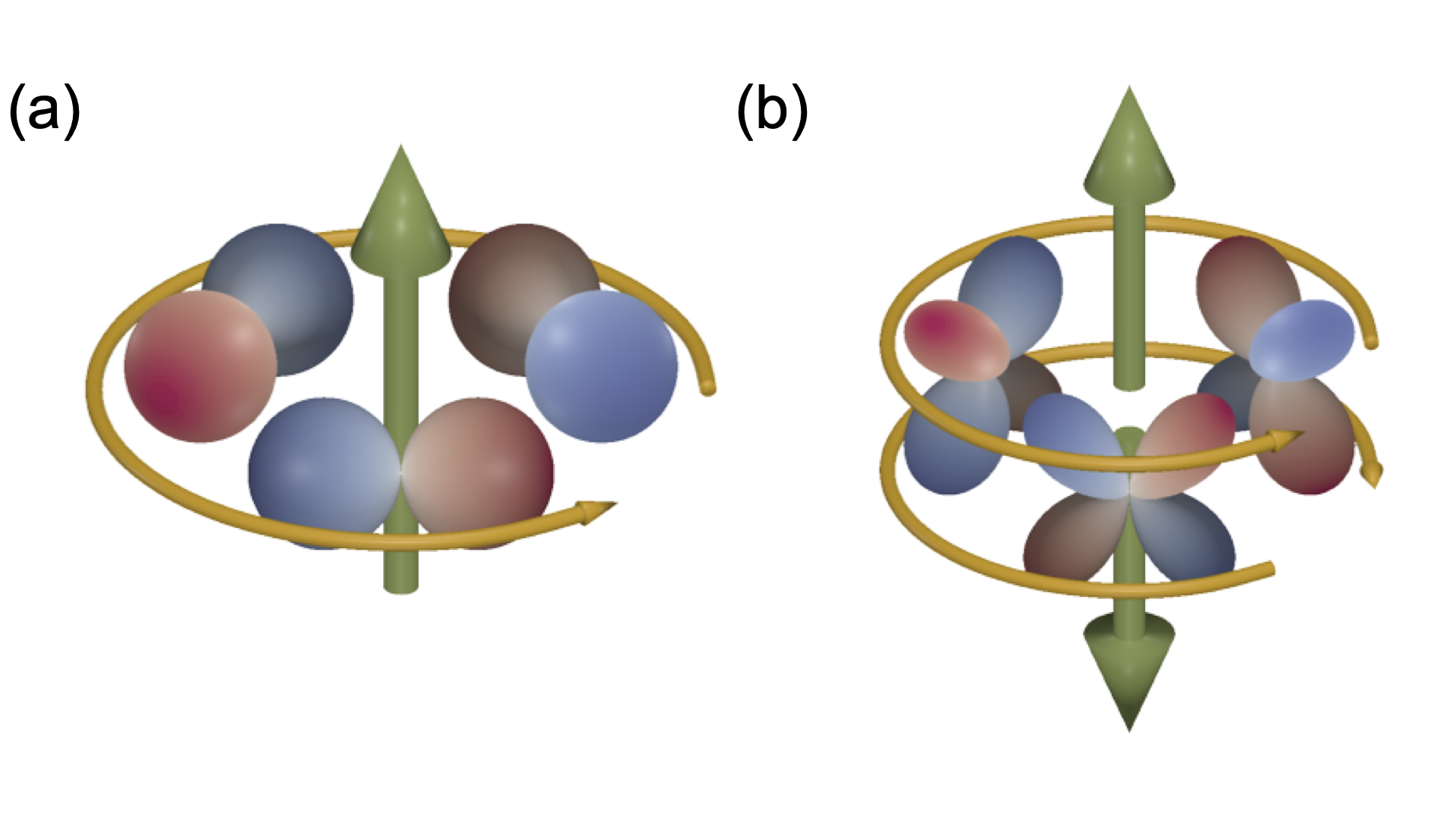}
\caption{\justifying
Schematic illustration of the intuitive images of (a) ETD and (b) ETQ moments under cluster-type orbital orderings. 
Red and blue indicate positive and negative phases, respectively, and green arrows denote the ETD. 
}
 \label{fig_G1G2ex}
\end{figure}

In many ferroaxial materials, the ferroaxial transition is characterized as a structural transition accompanied by rotational atomic displacements. 
However, studies of how such structural transitions affect the electronic degrees of freedom have been largely limited to model calculations~\cite{Inda_PhysRevB.111.L041104}, and quantitative analyses have been scarce~\cite{Bhowal_PhysRevResearch.6.043141}. 
Moreover, although several microscopic mechanisms have been proposed for the electric toroidal dipole (ETD)---the order parameter of ferroaxial order---such as vortices of electric polarization [Fig.~\ref{fig_G1G2ex}(a)]~\cite{dubovik1975multipole, dubovik1986axial} and cross-product--type spin--orbit coupling {(SOC)} $\bm{l} \times {\bm{\sigma}}$~\cite{Hayami_PhysRevB.98.165110}, where $\bm{l}$ and ${\bm{\sigma}}$ denote the orbital and spin angular momenta, their relative importance in real materials remains unclear. 
Furthermore, recent studies have shown that cluster-type orbital ordering also leads to a net ferroaxial moment corresponding to ETD [Fig.~\ref{fig_G1G2ex}(a)]~\cite{hayami2023cluster}, as well as an antiferro-type ferroaxial moment corresponding to an electric toroidal quadrupole (ETQ) [Fig.~\ref{fig_G1G2ex}(b)]~\cite{oiwa2025predominant, ishitobi2025purely}. 
Consequently, it is still unknown which electronic degrees of freedom play the primary role in characterizing ferroaxial order. 
These issues highlight the necessity of theoretical analyses that start from realistic electronic structures in materials.

\begin{figure}
\begin{center}
 \includegraphics[width=1\hsize]{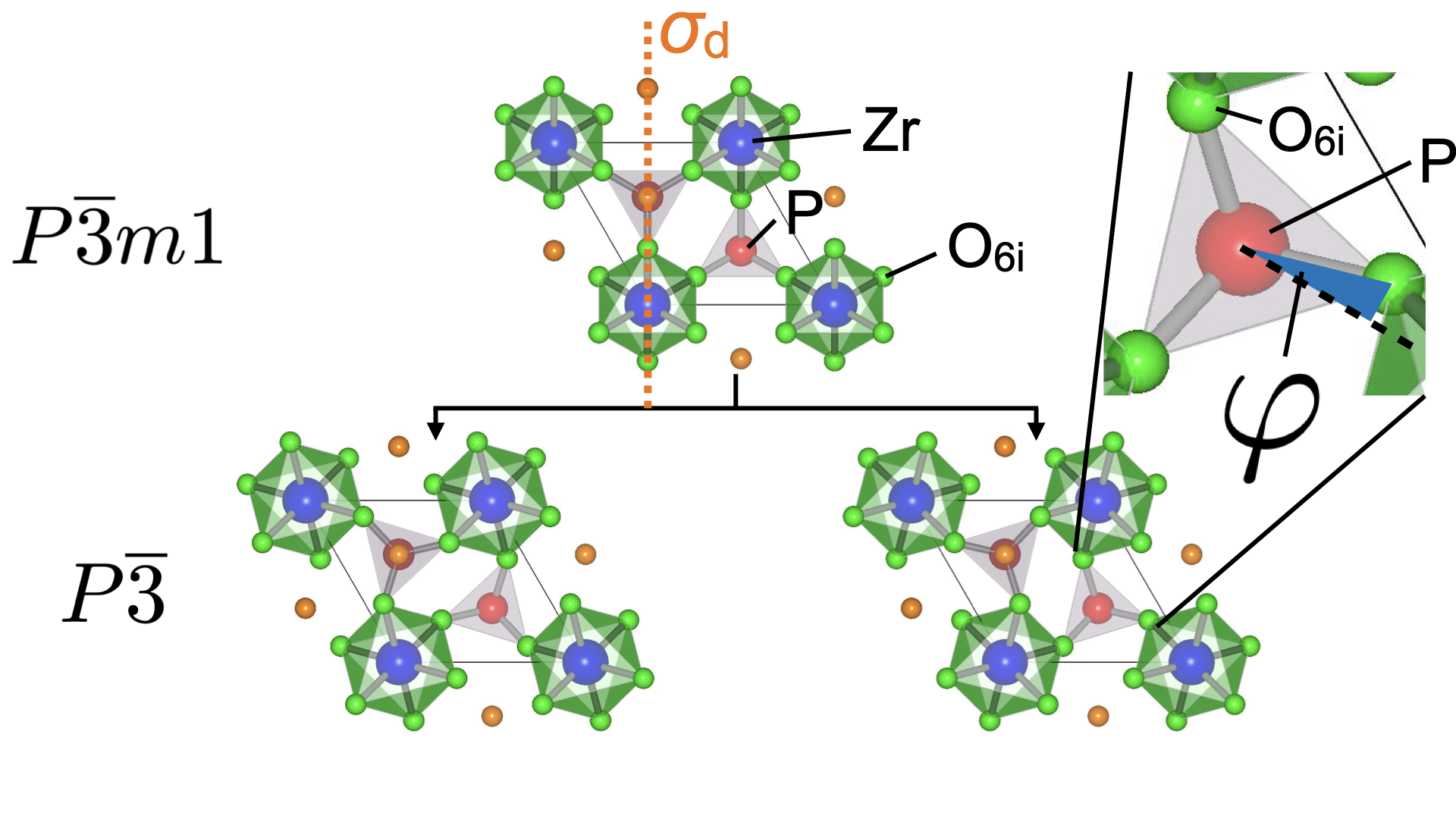}
\caption{\justifying
Crystal structure of ${\rm K_2Zr(PO_4)_2}$ in the nonferroaxial phase ($P\overline{3}m1$, top) and the ferroaxial phase ($P\overline{3}$, bottom). 
Atoms in blue, red, green, and orange correspond to Zr, P, ${\rm O_{6i}}$, and ${\rm O_{2d}}$ atoms, respectively, while K atoms are positioned underneath the P and ${\rm O_{2d}}$. 
Here,  ${\rm O_{6i}}$ and ${\rm O_{2d}}$ denote the O atoms at the 6i and 2d Wyckoff positions in $P\overline{3}m1$.
The displacement angle ${\varphi}$ of the ${\rm PO_4}$ tetrahedra characterizes the degree of ferroaxiality.
}
 \label{fig_structure}
 \end{center}
\end{figure}

Motivated by these considerations, we investigate the prototypical ferroaxial material, ${\rm K_2Zr(PO_4)_2}$, which undergoes a displacive-type ferroaxial transition at approximately 700~K~\cite{yamagishi2023ferroaxial}. 
In this transition, the crystal symmetry changes from the nonferroaxial structure $P\overline{3}m1$ (\#164, $D_{\rm 3d}^{1}$) to the ferroaxial structure $P\overline{3}$ (\#147, $C_{\rm 3i}^{1}$), as illustrated in Fig.~\ref{fig_structure}.
According to the rotation of the ${\rm PO_4}$ tetrahedra (or ${\rm ZrO_6}$ octahedra) by the displacement angle $\varphi$, the vertical mirror symmetry $\sigma_{\rm d}$ is broken while retaining threefold rotational symmetry.
Since $\varphi$ takes a finite value only in the $P\overline{3}$ phase, it serves as a quantitative measure of ferroaxiality from the viewpoint of the lattice degree of freedom.
In this situation, it is expected that corresponding electronic ferroaxial degrees of freedom also become nonzero. 
For example, it has been confirmed for the cross--product--type {SOC} defined at the P site~\cite{Bhowal_PhysRevResearch.6.043141}, whereas systematic quantitative comparisons among the various ferroaxial electronic degrees of freedom, including spinless degrees of freedom, have not yet been performed.

In the present study, we perform a detailed quantitative analysis of the ferroaxial compound ${\rm K_2Zr(PO_4)_2}$ to clarify the role of the microscopic electronic degrees of freedom under a static rotational distortion.
Employing density-functional theory (DFT) calculations in combination with a symmetry-adapted closest Wannier (SymCW) method~\cite{Ozaki_CW_PRB_2024, oiwa2025symmetry}, we examine how the ferroaxial-related electronic degrees of freedom evolve as a function of the displacement angle. 
Within a complete multipole basis set, we find that various multipolar degrees of freedom, such as the ETD, electric toroidal octupole (ETO), and electric hexadecapole (EH), emerge as multi-site spinless bases that significantly affect the electronic structure. 
Especially, we demonstrate that the {two} ETOs, expressed in terms of hybridizations between the $p$ orbitals on P atoms and $d$ orbitals on Zr atoms and between the $p$ orbitals on P and O atoms, provide the dominant contributions to the ferroaxial transition. 
We further show that relativistic {SOC} has only a negligible influence on the ferroaxial transition.

The remainder of this paper is organized as follows.  
In Sec.~\ref{sec_Allowedsambs}, we show the electronic degrees of freedom related to ferroaxiality in the displacive-type ferroaxial transition of ${\rm K_2Zr(PO_4)_2}$, based on the symmetry-adapted multipole framework.  
Section~\ref{sec_dftres} presents the electronic band structures and orbital-resolved density of states obtained by the DFT calculations. 
Section~\ref{sec_SCW} provides the effective tight-binding model {derived} from the DFT calculations using the SymCW method. 
Section~\ref{sec_mainres} presents a multipole-based classification of ferroaxial degrees of freedom, which allows us to identify the essential electronic degrees of freedom arising from the entanglement of orbital and bond.
We also present the dominant multipole contributions that emerge when the ferroaxial order occurs. 
Finally, Sec.~\ref{sec_conc} summarizes our findings and presents the perspective for future studies.

\section{Ferroaxial degrees of freedom based on multipole representation\label{sec_Allowedsambs}}
Ferroaxial degrees of freedom can be described unambiguously within the framework of the symmetry-adapted multipole basis (SAMB) theory, which has been established recently~\cite{Kusunose_PhysRevB.107.195118}. 
SAMBs consist of four types of multipoles, which are classified as electric (E: $Q$), magnetic (M: $M$), magnetic toroidal (MT: $T$), and electric toroidal (ET: $G$) multipoles according to their $\mathcal{P}$ and $\mathcal{T}$ parities~\cite{Hayami_PhysRevB.98.165110, kusunose2022generalization, hayami2024unified}.
For a given symmetry group, each multipole $X_{l\gamma}^{\Gamma}$ ($X=Q,M,T,G$) is further characterized by its rank $l$, the irreducible representation (IR) $\Gamma$, and its component $\gamma$. 
Since the SAMB constitutes a complete basis set in a given physical space, any degrees of freedom, such as orbital, spin, bond, and their composites, can be described in a unified way. 

\begin{table}
\setlength{\tabcolsep}{12pt}
\caption{\justifying
Correspondence between the atomic orbital hybridization and the atomic SAMBs {in ${\rm K_2Zr(PO_4)_2}$} without the spin degree of freedom.
The subscript associated with a multipole $X = Q, M, T, G$ specifies its rank.
}
\label{tab_AllowedAS}
\renewcommand{\arraystretch}{1.5}
\begin{tabular}{ccc}
\hline\hline
Orbital & \multicolumn{2}{c}{Atomic SAMBs} \\
 &\ $\mathcal{P}$-even\ & $\mathcal{P}$-odd \\
 \hline
$s$-$s$ & $Q_0$ & - \\
$s$-$p$ & - & $Q_1,\, T_1$\\
$s$-$d$ & $Q_2,\, T_2$ &-\\
$p$-$p$ & $Q_0,\, Q_2$,\, $M_1$ & -\\
$p$-$d$ & - & $Q_1,\, Q_3,\, G_2,$\\
 & &$T_1,\, T_3,\, M_2$\\
$d$-$d$ & $Q_0,\, Q_2,\, Q_4,$ & -\\
		      & $M_1,\, M_3$    &  \\
\hline\hline
\end{tabular}
\setlength{\tabcolsep}{6pt}
\end{table} 

\begin{figure*}[t]
\includegraphics[width=0.3\textwidth]{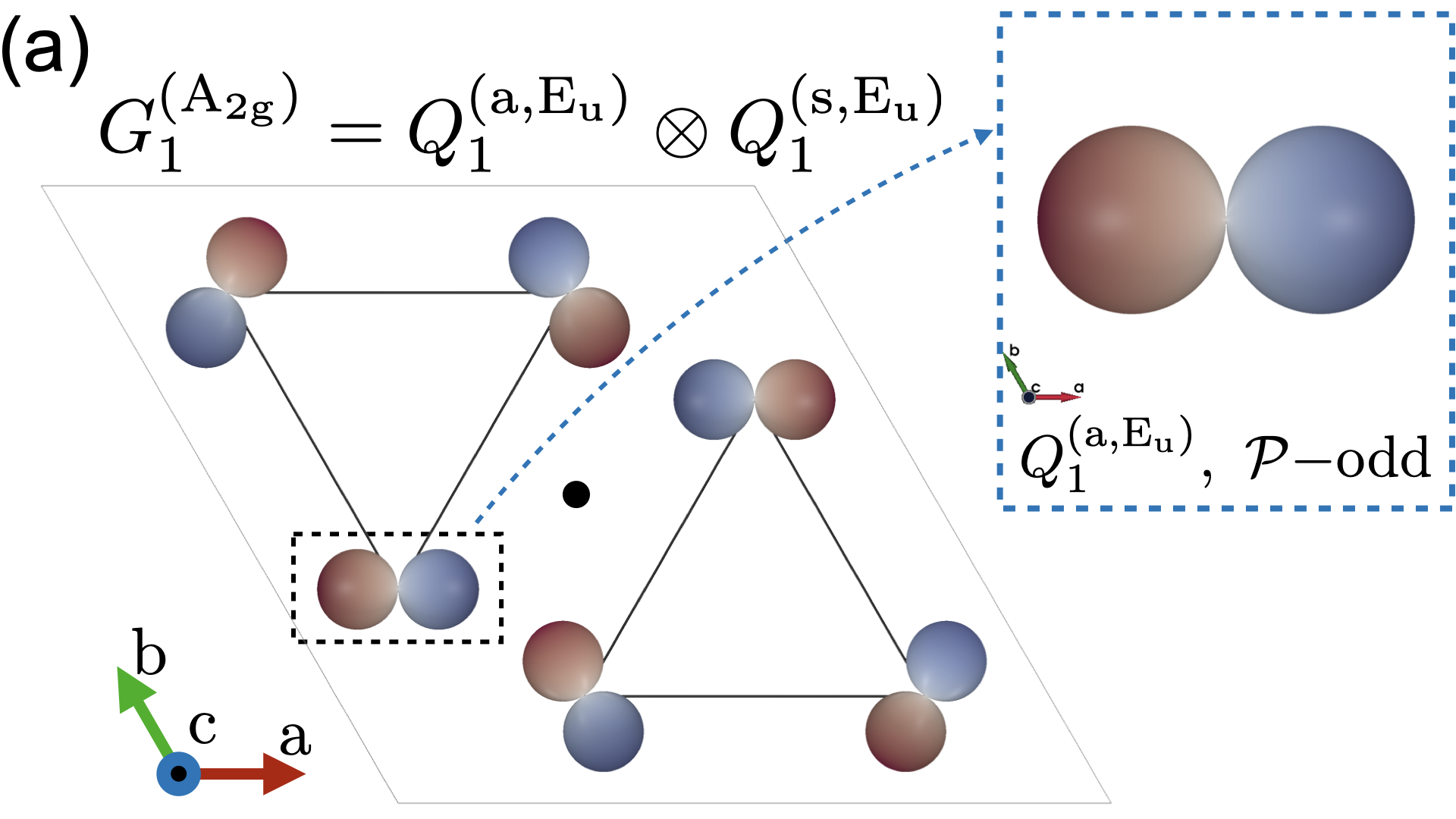} \includegraphics[width=0.3\textwidth]{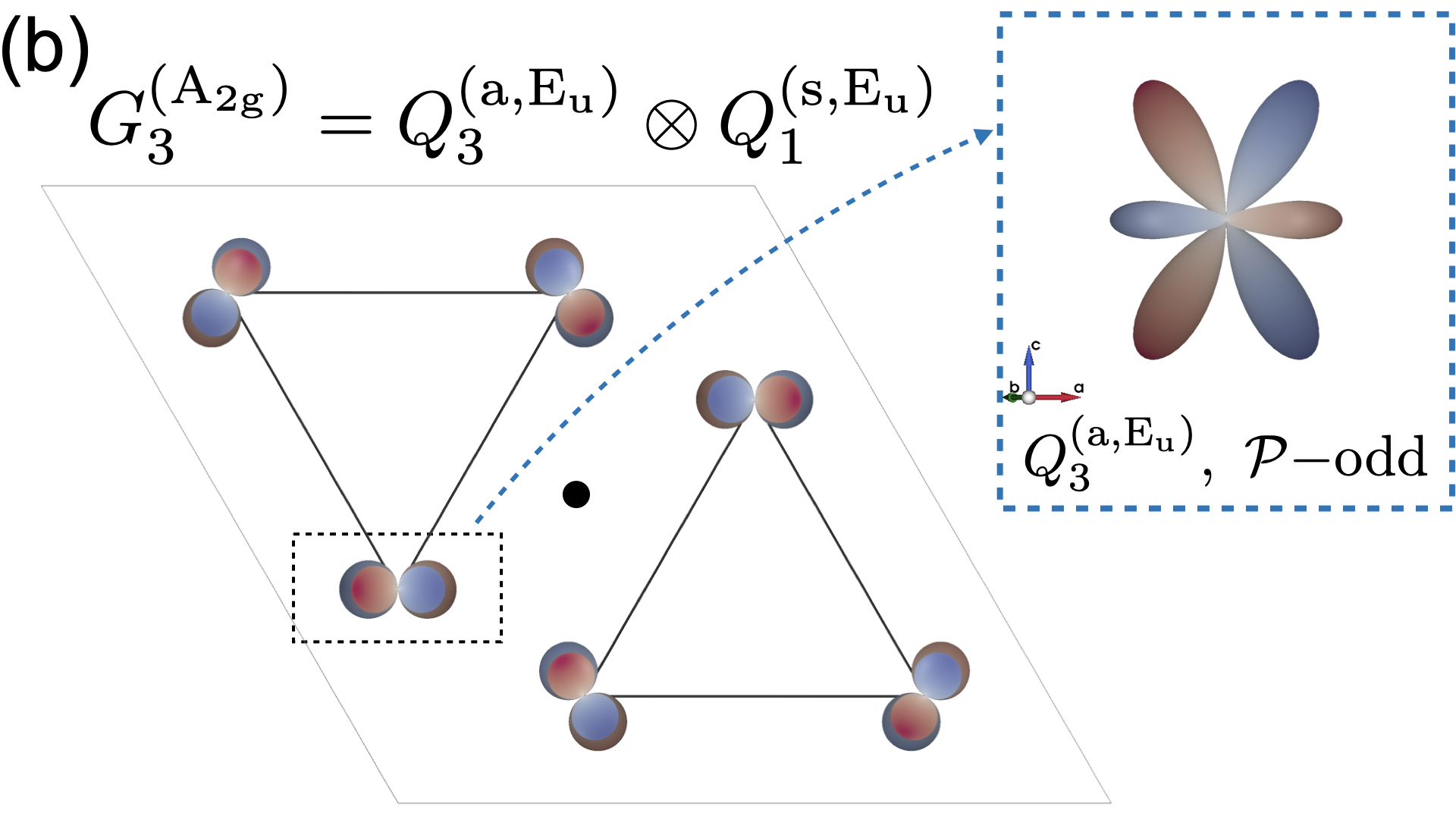} \includegraphics[width=0.3\textwidth]{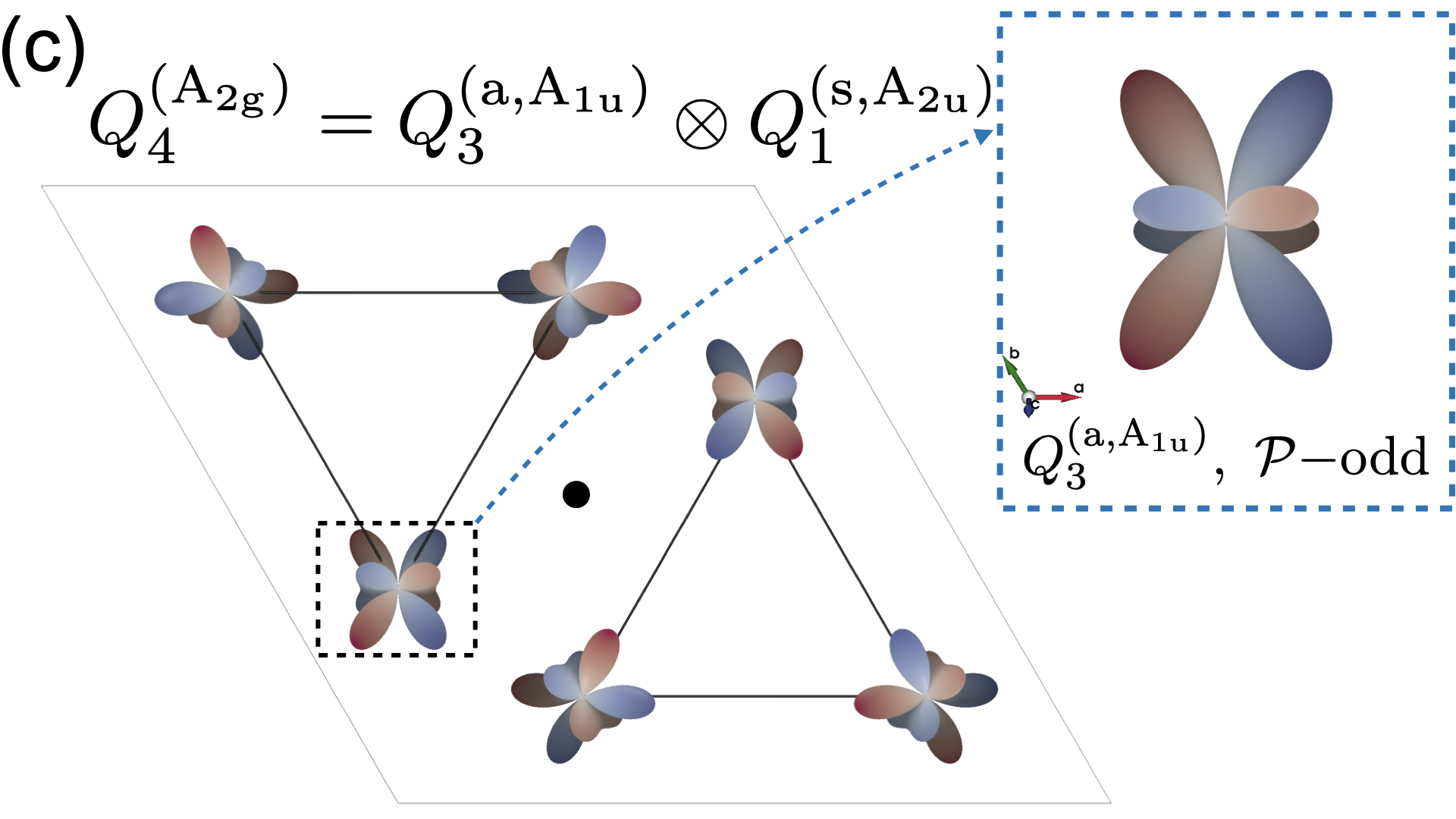}
\includegraphics[width=0.3\textwidth]{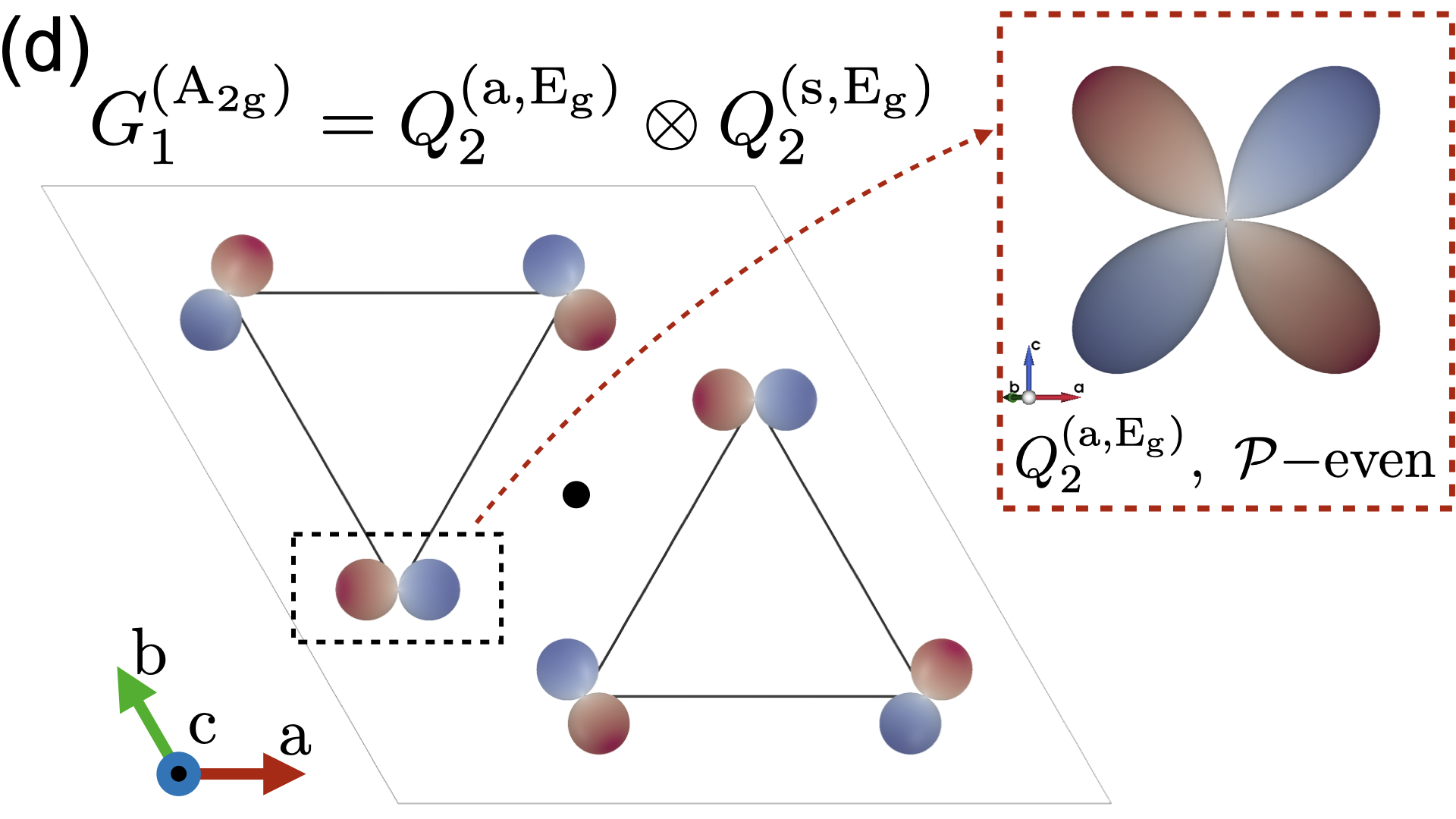} \includegraphics[width=0.3\textwidth]{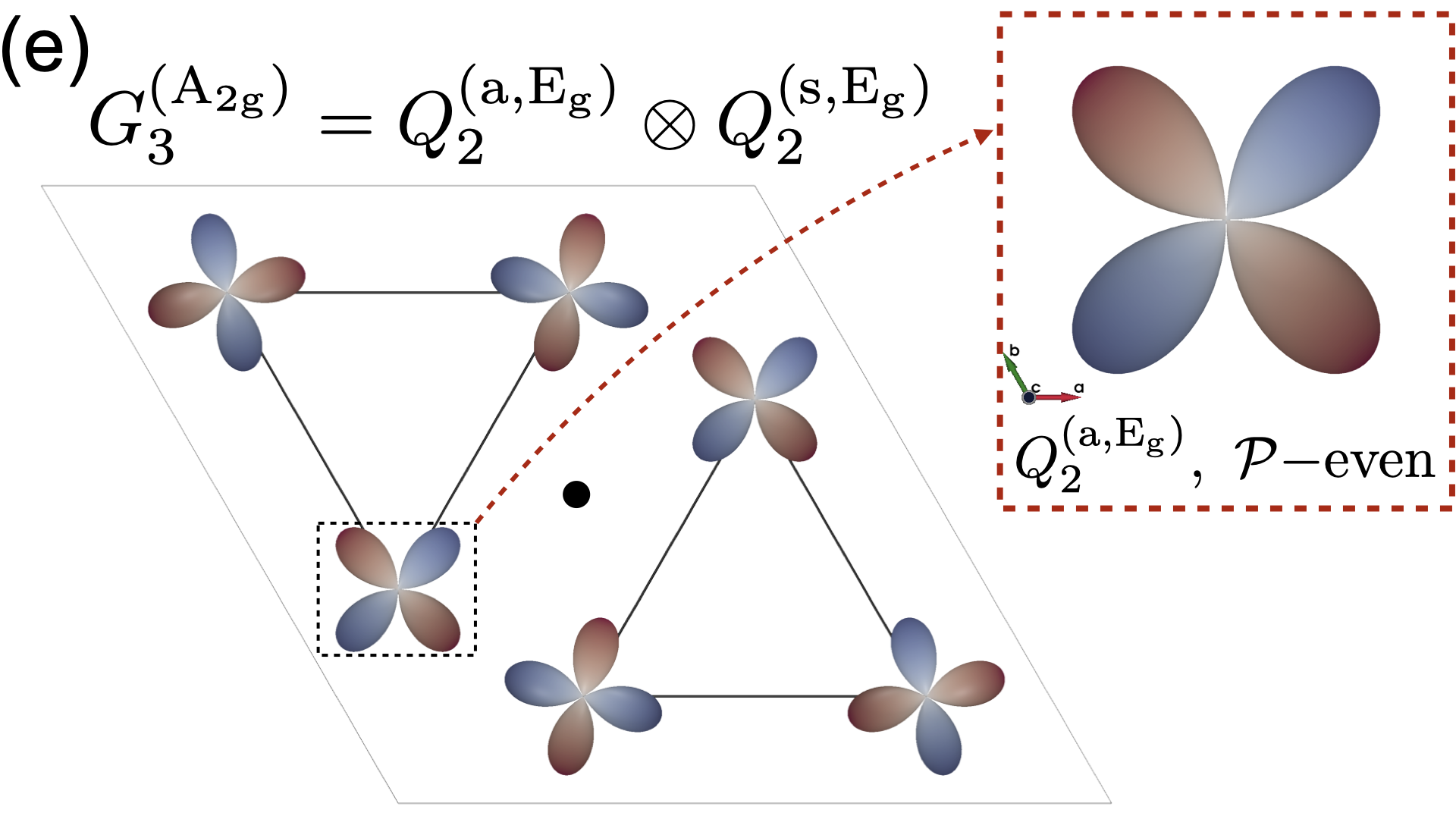} \includegraphics[width=0.3\textwidth]{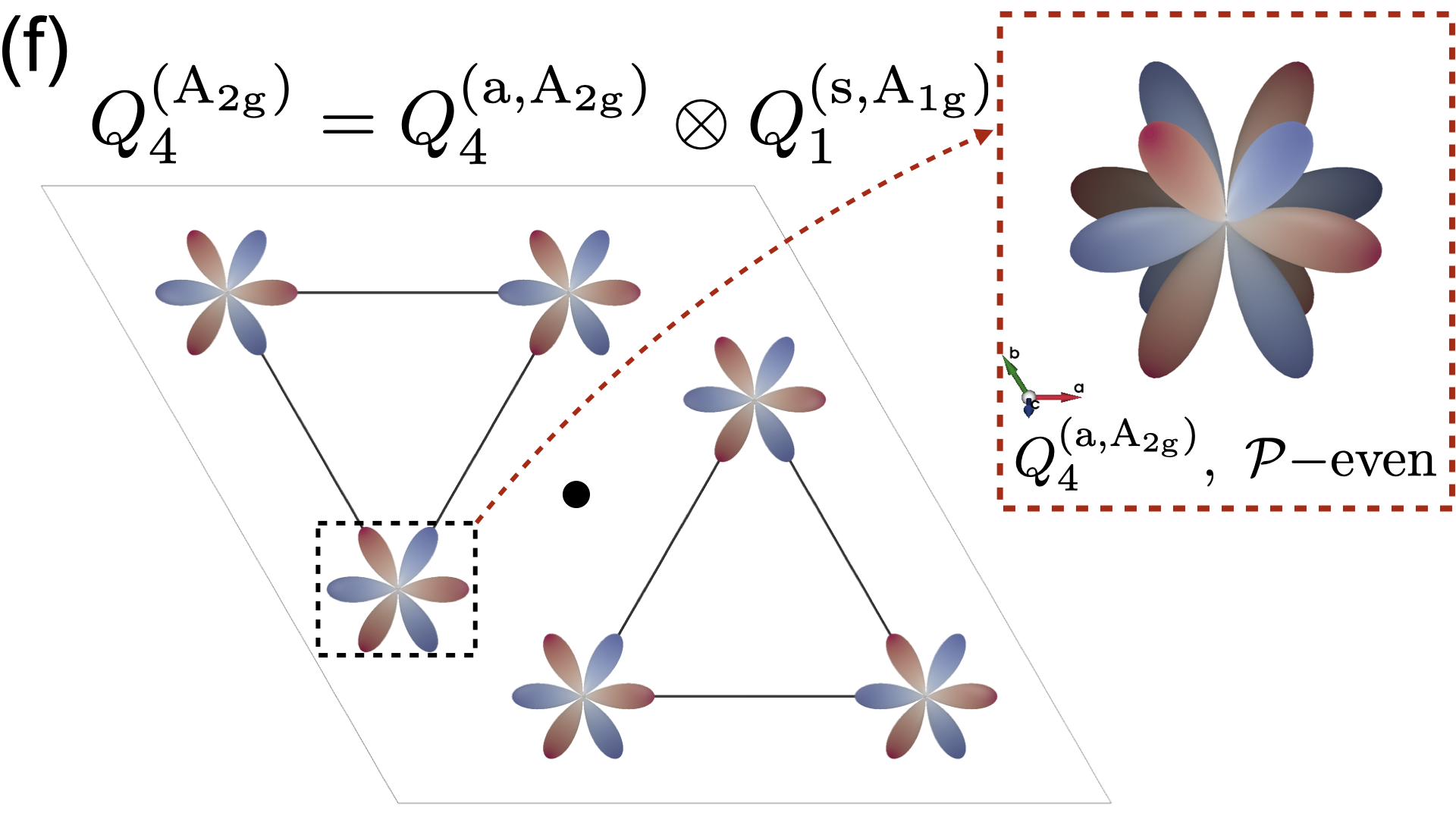}
\caption{\justifying
Examples of SAMBs belonging to the ${\rm A_{2g}}$ irreducible representation (IR) of the $D_{\rm 3d}$ point group in a six-site cluster.
The resulting ${\rm A_{2g}}$ multipoles are (a, d) the ETD, (b, e) the ETO, and (c,f) the EH. 
{The atomic SAMBs are $\mathcal{P}$-odd in (a)--(c) and $\mathcal{P}$-even in (d)--(f).}
All the SAMBs retain the spatial inversion symmetry in terms of the inversion center denoted by the black dots.
}
\label{fig_ExOfMultipoles}
\end{figure*}

For later convenience, we show the representative electronic ferroaxial degrees of freedom in ${\rm K_2Zr(PO_4)_2}$. 
We here consider the unit consisting of upward and downward triangles connected by the spatial inversion symmetry in Fig.~\ref{fig_ExOfMultipoles}, which corresponds to the PO$_4$ tetrahedra in Fig.~\ref{fig_structure}. 
Since the ferroaxial degree of freedom corresponds to the ${\rm A_{2g}}$ IR in the high-temperature nonferroaxial phase under the symmetry $P\overline{3}m1$, we show how to construct the electronic ferroaxial degrees of freedom belonging to the ${\rm A_{2g}}$ IR based on the SAMB.

The SAMB is constructed by the product of the atomic degrees of freedom and the site/bond degrees of freedom; the former is referred to as the atomic SAMB (denoted by $Y_{l}^{({\rm a},\Gamma)}$)~\cite{kusunose2020complete} and the latter is referred to as the cluster SAMB (denoted by $Y_{l}^{({\rm s},\Gamma)}$ for site and $Y_{l}^{({\rm b},\Gamma)}$ for bond) for $Y=Q,M,T,G$~\cite{Suzuki_PhysRevB.95.094406, Hayami_PhysRevLett.122.147602}. 
The atomic SAMBs relevant to ${\rm K_2Zr(PO_4)_2}$, spanning the $s$-, $p$-, and $d$-orbital manifolds, are summarized in Table~\ref{tab_AllowedAS}, where the $\mathcal{P}$-even multipoles ($Q_0$, $Q_2$, $Q_4$, $M_1$, $M_3$, and $T_2$) are activated in the orbital space with even-parity hybridization, while the $\mathcal{P}$-odd multipoles ($Q_1$, $Q_3$, $G_2$, $M_2$, $T_1$, and $T_3$) are activated in the orbital space with odd-parity hybridization. 
The cluster SAMBs relevant to ${\rm K_2Zr(PO_4)_2}$ can be obtained in an analogous manner by incorporating the permutation symmetries among the sites~\cite{Kusunose_PhysRevB.107.195118}.

Since the ${\rm A_{2g}}$ representation can be constructed from the direct products ${\rm A_{1g}}\otimes{\rm A_{2g}}$, ${\rm A_{1u}}\otimes{\rm A_{2u}}$, ${\rm E_{g}}\otimes{\rm E_{g}}$, and ${\rm E_{u}}\otimes{\rm E_{u}}$, the SAMBs corresponding to the ferroaxial degree of freedom are obtained by appropriately combining atomic and cluster SAMBs in a way that preserves the symmetry. 
We show several specific examples in the case of the site-cluster SAMB in Fig.~\ref{fig_ExOfMultipoles}. 
The ferroaxial degrees of freedom can be constructed by arranging the atomic electric dipole [Fig.~\ref{fig_ExOfMultipoles}(a)] and two different electric octupoles [Figs.~\ref{fig_ExOfMultipoles}(b) and \ref{fig_ExOfMultipoles}(c)], two different electric quadrupoles [Figs.~\ref{fig_ExOfMultipoles}(d) and \ref{fig_ExOfMultipoles}(e)], and electric hexadecaple [Fig.~\ref{fig_ExOfMultipoles}(f)] so as to satisfy spatial inversion and threefold rotational symmetries. 
By taking into account the rank of multipoles, the ferroaxial degrees of freedom in Figs.~\ref{fig_ExOfMultipoles}(a)--\ref{fig_ExOfMultipoles}(f) are identified as two ETDs [Figs.~\ref{fig_ExOfMultipoles}(a) and \ref{fig_ExOfMultipoles}(d)], two ETOs [Figs.~\ref{fig_ExOfMultipoles}(b) and \ref{fig_ExOfMultipoles}(e)], and two EHs [Figs.~\ref{fig_ExOfMultipoles}(c) and \ref{fig_ExOfMultipoles}(f)]. 
Similarly, one can construct the ferroaxial degrees of freedom by the product of atomic and bond-cluster SAMBs by considering the bond degree of freedom instead of the site one.

\section{DFT calculations} \label{sec_dftres}
We calculate the electronic states of ${\rm K_2Zr(PO_4)_2}$ based on DFT calculations by using Quantum ESPRESSO~\cite{Giannozzi_2009}.
In the following calculations, we adopt the PBE exchange-correlation functional and use the scalar relativistic optimized norm-conserving Vanderbilt (ONCV) pseudopotentials~\cite{ONCV_PRB_2013} downloaded from PseudoDojo \cite{VANSETTEN201839}. 
First, by performing optimization of atomic positions with the fixed lattice constants $a = 5.2759${~\AA~and $c = 9.0651$~\AA}, we obtain the optimal atomic displacement angle ${\varphi} = \pm 16.89^\circ$, which is consistent with the experimental value of $\varphi \sim 17^{\circ}$ at 300 K~\cite{yamagishi2023ferroaxial}. Here the positive (negative) sign of ${\varphi}$ is defined as counterclockwise (clockwise) rotations of the ${\rm PO_4}$ tetrahedra.

\begin{figure}
 \includegraphics[width=0.95\linewidth]{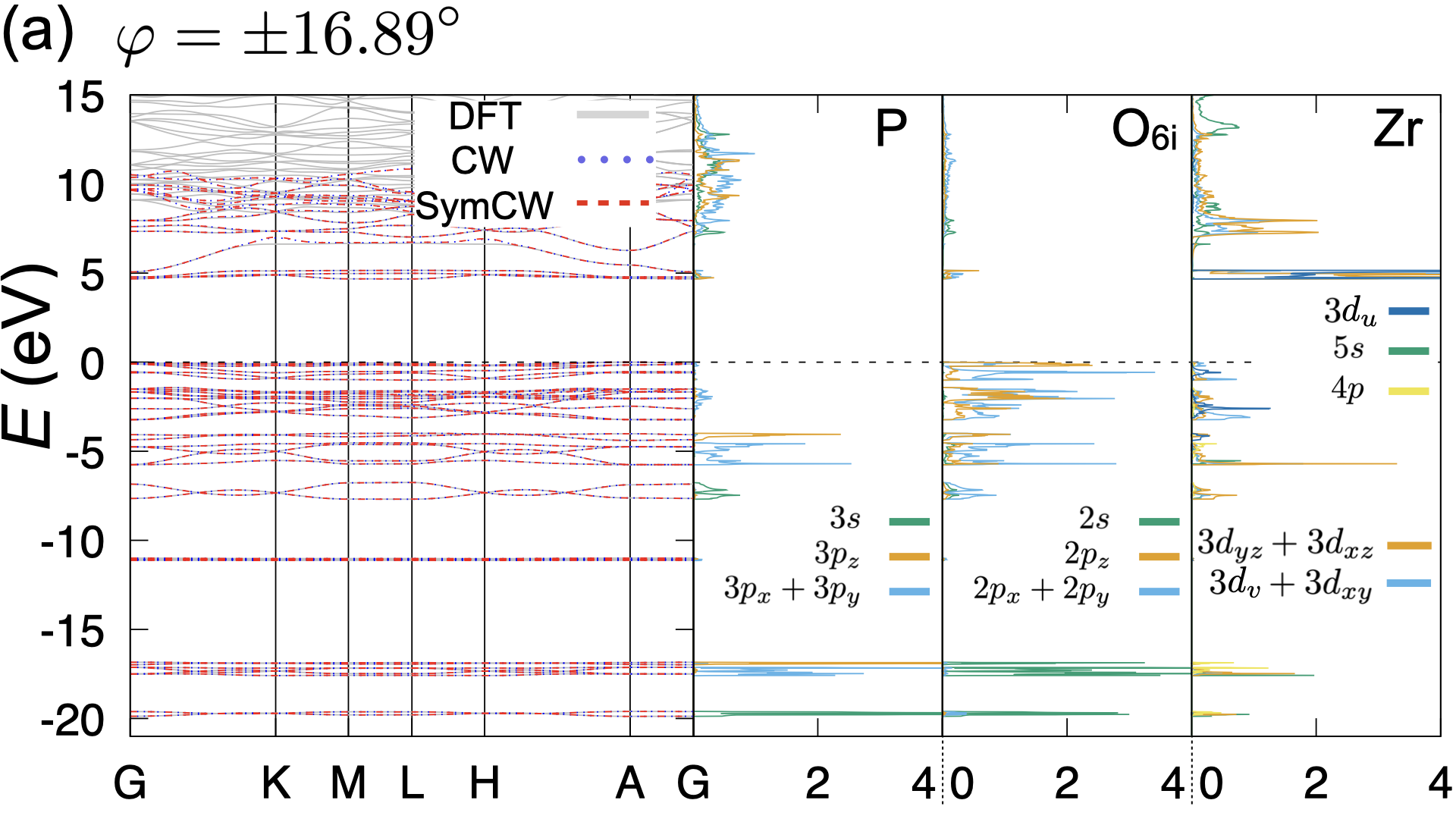} \includegraphics[width=0.95\linewidth]{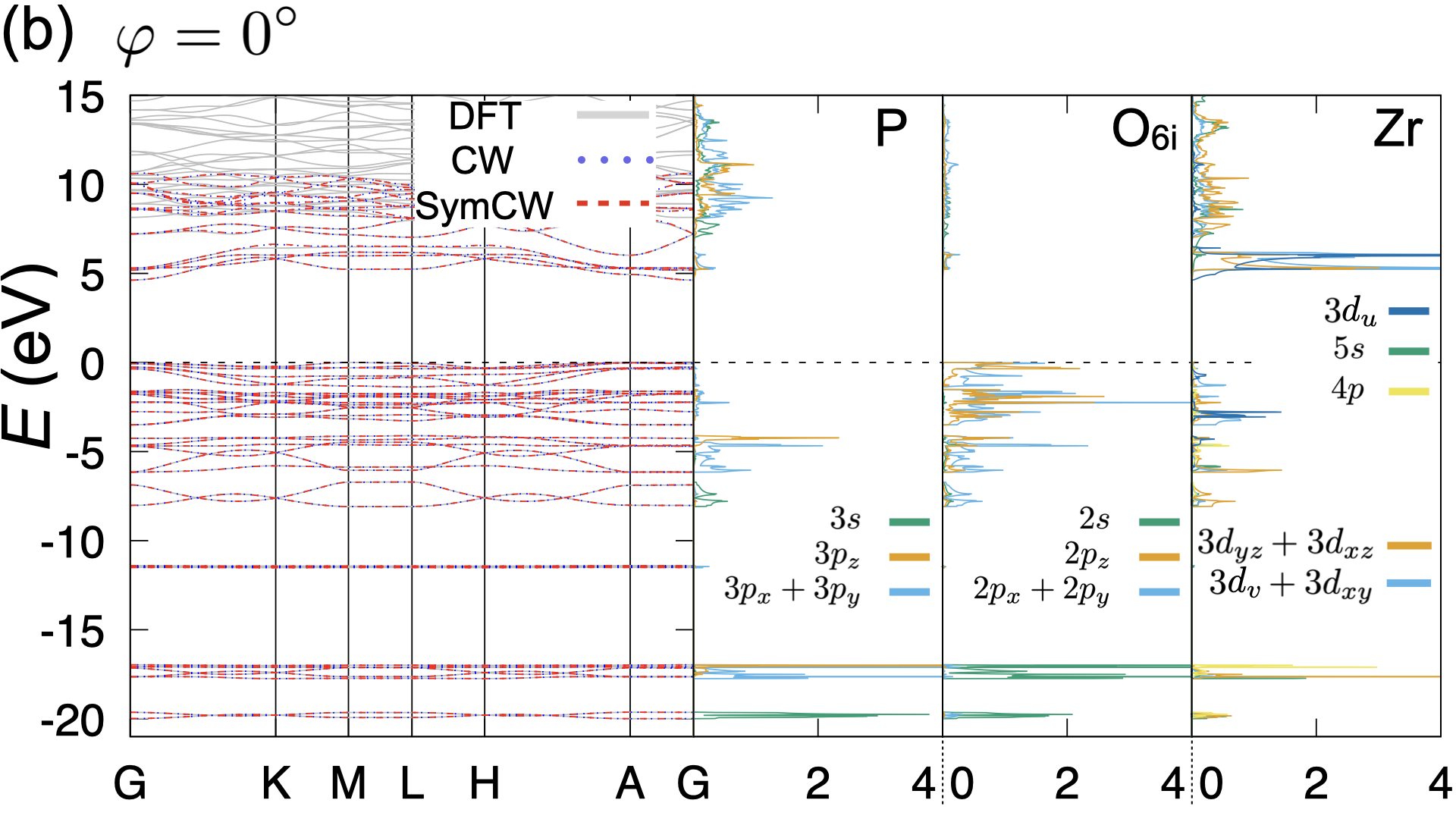}
\caption{\justifying
Band structures and densities of states (DOS) obtained from DFT calculations, together with the fitting results of the CW and SymCW models. Panels (a) and (b) correspond to displacement angles of $\varphi = \pm16.89^\circ$ and $\varphi = 0^\circ$, respectively. 
{In the left panel, g}ray lines represent the DFT {bands}, while red and blue lines show {the band structures obtained by} the CW and SymCW model{s}. 
The DOS per atom for orbitals of P, ${\rm O_{6i}}$, and Zr are shown {in the right panel}; the in-plane ($xy$) {and $d_{u} \equiv d_{3z^2-r^2}$} component{s are} represented in blue, the out-of-plane component in orange{, the $s$ orbital in green{, and the $4p$ orbital in yellow}.}
The proposed models reproduce the DFT results well up to about 7 eV above the Fermi level, which is the energy range relevant to most physical properties.}
 \label{fig_bandDOS}
\end{figure}

Then, we perform self-consistent DFT calculations for $\pm 16.89^\circ$ and ${\varphi} = 0^\circ$. 
The kinetic-energy cutoff for the Kohn--Sham (KS) orbitals and the total--energy convergence threshold are set to 100 Ry and $1\times10^{-14}$ Ry, respectively.
The $\boldsymbol{k}$-point grid is set to $12\times12\times6$.
The obtained band structures and partial (orbital-resolved) density of states (DOS) for  ${\varphi} =\pm 16.89^\circ$ and $0^\circ$ are shown in Figs.~\ref{fig_bandDOS}(a) and (b), respectively.
According to the DOS analysis, the $s$ and $p$ orbitals of P and O$_{\rm 6i}$, and the $d$ orbitals of Zr contribute predominantly to the electronic states around $-20~{\rm eV} \lesssim E \lesssim 7~{\rm eV}$, including the Fermi level{, which is set to 0~eV}.
In contrast, the $s$ and $p$ orbitals of Zr make negligible contributions near the Fermi level{.} The contributions of the $4s$ and $3p$ orbitals of K, as well as the $2s$ and $2p$ orbitals of O$_{\rm 2d}$ {{are {also} finite},} which are not shown in Fig.~\ref{fig_bandDOS}.
Here and hereafter, we denote two O atoms at the 6i and 2d Wyckoff positions in $P\overline{3}m1$ as ${\rm O_{6i}}$ and ${\rm O_{2d}}$.

\section{SYMMETRY-ADAPTED CLOSEST WANNIER MODEL \label{sec_SCW}}

In this section, we construct the symmetry-adapted closest Wannier model for ${\rm K_2Zr(PO_4)_2}$ by using the SymCW method, implemented in the open-source Python library SymClosestWannier~\cite{Ozaki_CW_PRB_2024, oiwa2025symmetry}.

For the initial guess of the closest Wannier functions (CWFs), we use the hydrogenic atomic $s$ and $p$ orbitals for P, K, O$_{\rm 6i}$, O$_{\rm 2d}$ {and $d$ orbitals for Zr}, where the total number of orbitals is 53 {in the unit cell}. The $s$ and $p$ orbitals of Zr are omitted from the model because their contributions are negligible.
Note that the symmetry properties of the CWFs are common to those of the original atomic orbitals used for the initial guess~\cite{Ozaki_CW_PRB_2024, oiwa2025symmetry}.
Therefore, the SAMBs can be defined as a complete orthonormal basis set $\{ \mathbb{Z}_j \}$ of the CWFs, which allows the $53\times53$ closest Wannier (CW) tight-binding Hamiltonian, $H^{\rm CW}$, to be expressed as a linear combination of $\{ \mathbb{Z}_j \}$:
\begin{align}
&H^{\rm SymCW} =\sum_j^{\Gamma_j \in {\rm A_{1g},\,{A_{2g}}}}z_j\mathbb{Z}_j, 
\label{eq_H_SymCW}
\\
&z_j=\mathrm{Tr}\left(\mathbb{Z}_jH^{\rm CW}\right){.}
\label{eq_zj}
\end{align}
Here, we ignore the spin degree of freedom, which will be discussed in Sec.{~\ref{sec: Effect of relativistic spin–orbit coupling}}.
We refer to this symmetrized model as the SymCW model{.} The SAMBs $\{\mathbb{Z}_j\}$, belonging to the ${\rm A_{1g}}$ and ${\rm A_{2g}}$ IRs of the $D_{\rm 3d}$ point group, are automatically generated using the open-source Python library MultiPie \cite{Kusunose_PhysRevB.107.195118}.
Here $z_j$ represents the weight of each SAMB $\mathbb{Z}_j$ with the {unit} of energy.
The total number of SAMBs is 96,439, by considering up to the 200 bond-clusters for each bond (the maximum bond length is about the distance between the home unit-cell and the 10th neighbor unit-cell).
The mean absolute error between the energy eigenvalues obtained from the DFT calculations and the SymCW models under the Fermi level is evaluated as
\begin{equation}
\eta =
\sum_{\boldsymbol{k}}
\sum_{n = 1}^{N_{\rm F}}
\frac{
\left|
  \epsilon_{n\boldsymbol{k}}^{\rm DFT}
  -
  \epsilon_{n\boldsymbol{k}}^{\rm SymCW}
\right|
}
{N_k N_{\rm F}}=
\begin{cases}
0.02667\; {\rm meV} \\
\hspace{1cm} (\varphi=\pm16.89^\circ),\\
0.02238\; {\rm meV} \\
\hspace{1cm} \hspace{25pt}(\varphi=0^\circ),\\
\end{cases}
\end{equation}
where $N_{\rm F} $ denotes the number of eigenvalues below the Fermi level and $N_k = 864~(12\times12\times6)$ is the number of $\boldsymbol{k}$ points used in the DFT calculations.
As shown in Fig.~\ref{fig_bandDOS}, the CW and SymCW models accurately reproduce the valence band structure of DFT calculations for both the $P\overline{3}$ ($\varphi = \pm 16.89^\circ$) and $P\overline{3}m1$ ($\varphi = 0^\circ$) phases.
Although the conduction bands above 7 eV, particularly those near the K and H points, are not well reproduced, this inaccuracy does not affect the results discussed below.

\section{RESULTS AND DISCUSSIONS \label{sec_mainres}}

In this section, we present an analysis on the spinless electric toroidal multipoles in the ferroaxial phase of ${\rm K_2Zr(PO_4)_2}$, which affects the electronic structure.
We first show the explicit definitions of several SAMBs together with their physical interpretations, highlighting their roles as key microscopic ingredients of the ferroaxial order.
We then present the evolution of their expectation value of each SAMB with varying the angle $\varphi$ and identify the dominant contributions to the ferroaxial transition.

\subsection{Microscopic representation of ferroaxial mutlipoles}
Let us first show the microscopic representation of SAMBs describing the ferroaxial order and present the results of an analysis of our models for $P\overline{3}$ ($\varphi = \pm 16.89^\circ$) and $P\overline{3}m1$ ($\varphi = 0^\circ$) structures.
The SymCW Hamiltonian in Eq.~(\ref{eq_H_SymCW}) can be rewritten as
\begin{align}
H^{\rm SymCW} &= H_{\rm CEF} + H_{\rm hop}, 
\label{eq_H_SymCW_2} \\
H_{\rm CEF} &= H_{\rm CEF}^{\rm (K)} + H_{\rm CEF}^{\rm (Zr)} + H_{\rm CEF}^{\rm (P)} + H_{\rm CEF}^{\rm (O_{6i})} + H_{\rm CEF}^{\rm (O_{2d})} , \\
H_{\rm hop} &= \sum_{n=1}^{200} \sum_{\Braket{M, N}}^{\rm Zr, K, P, O_{\rm 6i}, O_{\rm 2d}} H_{\rm hop}^{(M-N, n)}, 
\end{align}
where $H_{\rm CEF}^{(M)}$ is the crystalline electric field (CEF) term at $M =$ K, Zr, P, O$_{\rm 6i}$, O$_{\rm 2d}$ atoms and $H_{\rm hop}^{(M-N, n)}$ represents the spin-independent hopping term in the $n$-th bond-cluster of the $M$--$N$ bond between $M$ and $N$ atoms.
Each term is expressed as the linear combination of SAMBs $\{ \mathbb{Z}_j \}$, and their coefficients $\{ z_j \}$ can be obtained by computing Eq.~(\ref{eq_zj}).
We consider 96,439 SAMBs by including up to the 200 bond-clusters for each bond.
Among these SAMBs, {14121} is the number of multipoles belonging to the ${\rm A_{2g}}$ IR of the $D_{\rm 3d}$ point group symmetry.
Espeically, we hereafter focus on the representative ferroaxial SAMBs including ETD $G_{z}$, ETOs $G_{3b}$ and $G_{z\alpha}$, and EH $Q_{4b}$, whose coefficients are significantly large.
As for the hopping terms, we restrict our analysis to the nearest-neighbor term $H_{\rm hop}^{(M-N, 1)}$, since we have confirmed that the further-neighbor hoppings are negligibly small, with their magnitudes decaying exponentially beyond the second-nearest neighbors.

\begin{table*}
\caption{\justifying
Definitions of SAMBs belonging to the ${\rm A_{2g}}$ IR of the $D_{\rm 3d}$ point group symmetry for $P\overline{3}$ ($\varphi = \pm 16.89^\circ$) and $P\overline{3}m1$ ($\varphi = 0^\circ$) structures.
$z_j$ and $\braket{\mathbb{Z}_j}$ are the coefficient{s} of the SAMB given by Eq.~(\ref{eq_zj}) and the expectation value of the SAMB.
The symbols ${Q}^{\rm (a)}$ and ${Q}^{\rm (s/b)}$ represent the atomic {electric} and site- or bond-cluster {electric} bases, respectively. {The subscript of ${Q}^{\rm (a)}$ and ${Q}^{\rm (s/b)}$ represent the notation of {electric} bases, which are summarized at \cite{Hayami_PhysRevB.98.165110, Kusunose_PhysRevB.107.195118}.}
The column of ``Hybridization" represents the corresponding Hilbert space.
The angle $\varphi$ is defined in Fig.~\ref{fig_structure}.
}
\label{tab_ImportSambs}
\begin{centering}
\renewcommand{\arraystretch}{3.0}
\begin{tabular*}{\textwidth}{@{\extracolsep{\fill}} c c c c c}
\hline\hline
&
SAMB&Hybridization& $z_{j}$ [eV] & $\braket{\mathbb{Z}_j}$ \\
&&& $\makecell{\varphi = \pm 16.89^\circ  \\  (\varphi = 0^\circ)}$ 
& $\makecell{\varphi = \pm 16.89^\circ  \\  (\varphi = 0^\circ)}$ \\
\hline

$H^{\rm (O_{6i})}_{\rm CEF}$
& $\mathbb{Z}_{1}^{(G_{z})} = \frac{1}{\sqrt{2}}\left( Q_{x}^{({\rm a,\,E_{u}})}\otimes Q_{y}^{(\rm s, \, E_{u})} - Q_{y}^{(\rm a, \, E_{u})}\otimes Q_{x}^{({\rm s,\,E_{u}})} \right)$ & $\braket{s,\,{\rm O_{6i}}|p,\,{\rm O_{6i}}}$ & $\makecell{ \mp 2.61 \times 10^{-3} \\
( -4.68 \times 10^{-12}) }$ & $\makecell{ \mp 1.59 \times 10^{-1} \\ ( -2.45 \times 10^{-14}) }$ \\

& $\mathbb{Z}_{2}^{(G_{3b})} = \frac{1}{\sqrt{2}}\left( Q_{xy}^{(\rm a, \, E_{g})}\otimes Q_{yz}^{({\rm s,\,E_{g}})} - Q_{v}^{({\rm a,\,E_{g}})}\otimes Q_{zx}^{(\rm s, \, E_{g})} \right)$ & $\braket{p,\,{\rm O_{6i}}|p,\,{\rm O_{6i}}}$ & $\makecell{  \mp 8.90 \times 10^{-
1}  \\ (3.72 \times 10^{-11}) }$ & $\makecell{  \mp 8.11 \times 10^{-2}  \\ (-5.06 \times 10^{-12}) }$ \\

 \hline
 
 $H^{\rm (Zr)}_{\rm CEF}$ &  $\mathbb{Z}_{3}^{(Q_{4b})} = Q_{4b}^{({\rm a,\, A_{2g}})} \otimes Q_0^{({\rm s, \, A_{1g}})}$ & $\braket{d,\,{\rm Zr}|d,\,{\rm Zr}}$ & $\makecell{ \pm 3.79 \times 10^{-1}  \\ ( 6.01 \times 10^{-11} ) }$ & $\makecell{ \mp 8.74 \times 10^{-2}  \\ ( -1.72 \times 10^{-12} ) }$ \\

\hline

$H^{\rm (P-O_{6i}, 1)}_{\rm hop}$
 & $\mathbb{Z}_{4}^{(G_z)} = \frac{1}{\sqrt{2}}\left( Q_{x}^{({\rm a,\,E_{u}})} \otimes Q_{y}^{(\rm b, \, E_{u})} - Q_{y}^{(\rm a, \, E_{u})}\otimes Q_{x}^{({\rm b,\,E_{u}})}  \right)$  & $\braket{p,\,{\rm P}|s,\,{\rm O_{6i}}}$ & $\makecell{ \pm 7.10 
  \\ ( -9.73 \times 10^{-11}) }$ & $\makecell{ \mp 2.45 \times 10^{-
1}  \\ ( 5.09 \times 10^{-12}) }$ \\
 
 & $\mathbb{Z}_{5}^{(G_z)} = \frac{1}{\sqrt{2}}\left( Q_{x}^{({\rm a,\,E_{u}})} \otimes Q_{y}^{(\rm b, \, E_{u})} - Q_{y}^{(\rm a, \, E_{u})}\otimes Q_{x}^{({\rm b,\,E_{u}})}  \right)$  & $\braket{s,\,{\rm P}|p,\,{\rm O_{6i}}}$ & $\makecell{ \mp 3.54 
 \\ ( -1.26 \times 10^{-11}) }$ & $\makecell{ \pm 1.74 \times 10^{-
1}  \\ ( 7.41 \times 10^{-12}) }$ \\
 
 &  $\mathbb{Z}_{6}^{(G_z)} = \frac{1}{\sqrt{2}}\left( Q_{zx}^{(\rm a, \, E_{g})}\otimes Q_{yz}^{({\rm b,\,E_{g}})} - Q_{yz}^{({\rm a,\,E_{g}})}\otimes Q_{zx}^{(\rm b, \, E_{g})}  \right)$ & $\braket{p,\,{\rm P}|p,\,{\rm O_{6i}}}$ & $\makecell{ \pm 2.86 
 \\ ( 1.45 \times 10^{-10}) }$ & $\makecell{ \mp 1.73 \times 10^{-
1} \\ ( -1.96 \times 10^{-11}) }$ \\

 & $\mathbb{Z}_{7}^{(G_{3b})} = \frac{1}{\sqrt{2}}\left( Q_{xy}^{(\rm a, \, E_{g})}\otimes Q_{yz}^{({\rm b,\,E_{g}})} - Q_{v}^{({\rm a,\,E_{g}})}\otimes Q_{zx}^{(\rm b, \, E_{g})} \right)$ & $\braket{p,\,{\rm P}|p,\,{\rm O_{6i}}}$ & $\makecell{ \pm 7.21
 \\ ( 6.02 \times 10^{-12}) }$ & $\makecell{ \mp 4.60 \times 10^{-
1} \\ ( -4.26 \times 10^{-13}) }$ \\

 \hline

$H^{\rm (Zr-O_{6i}, 1)}_{\rm hop}$
& $\mathbb{Z}_{8}^{(G_{z})} = \frac{1}{\sqrt{2}}\left( Q_{zx}^{(\rm a, \, E_{g})}\otimes Q_{yz}^{({\rm b,\,E_{g}})} - Q_{yz}^{({\rm a,\,E_{g}})}\otimes Q_{zx}^{(\rm b, \, E_{g})} \right)$ & $\braket{d,\,{\rm Zr}|s,\,{\rm O_{6i}}}$ & $\makecell{   \pm 2.55 
  \\ (1.29 \times 10^{-10}) }$ & $\makecell{   \mp 1.37 \times 10^{-
1}  \\ (-1.69 \times 10^{-12}) }$ \\

 & $\mathbb{Z}_{9}^{(G_{3b})} = \frac{1}{\sqrt{2}}\left( Q_{xy}^{(\rm a, \, E_{g})}\otimes Q_{yz}^{({\rm b,\,E_{g}})} - Q_{v}^{({\rm a,\,E_{g}})}\otimes Q_{zx}^{(\rm b, \, E_{g})} \right)$ & $\braket{d,\,{\rm Zr}|s,\,{\rm O_{6i}}}$ & $\makecell{  \pm 3.50 
 \\ (-2.66 \times 10^{-11} ) }$ & $\makecell{  \mp 1.79 \times 10^{-
1}  \\ (-6.88 \times 10^{-13} ) }$ \\ 
 
 & $\mathbb{Z}_{10}^{(G_{z\alpha})} = \frac{1}{\sqrt{2}}\left( Q_{3u}^{({\rm a,\,E_{u}})}\otimes Q_{y}^{(\rm b, \, E_{u})}-Q_{3v}^{(\rm a, \, E_{u})}\otimes Q_{x}^{({\rm b,\,E_{u}})} \right)$ & $\braket{d,\,{\rm Zr}|p,\,{\rm O_{6i}}}$ & $\makecell{   \mp 8.70 \times 10^{-1}  \\ (-1.41 \times 10^{-12}  ) }$ & $\makecell{   \pm 1.02 \times 10^{-1}  \\ (-1.47 \times 10^{-12}  ) }$ \\ 
 
 & $\makecell{\mathbb{Z}_{11}^{(G_{z\alpha})} = \frac{1}{2\sqrt{2}} \left( Q_{z\beta}^{({\rm a,\,E_{u}})}\otimes Q_{x}^{(\rm b, \, E_{u})}-Q_{xyz}^{(\rm a, \, E_{u})}\otimes Q_{y}^{({\rm b,\,E_{u}})}  \right. \\
\hfill \left. - \sqrt{6} Q_{3a}^{({\rm a,\,A_{1u}})}\otimes Q_z^{({\rm b, A_{2u}})} \right)}$  & $\braket{d,\,{\rm Zr}|p,\,{\rm O_{6i}}}$ & $\makecell{  \pm 4.35 \\ (4.51 \times 10^{-11}   ) }$ & $\makecell{  \mp 3.80 \times 10^{-1}  \\ (-1.28 \times 10^{-12}   ) }$ \\

 & $\makecell{\mathbb{Z}_{12}^{(Q_{4b})} = \frac{3}{2\sqrt{6}} \left(Q_{z\beta}^{({\rm a,\,E_{u}})}\otimes Q_{x}^{(\rm b, \, E_{u})}- Q_{xyz}^{(\rm a, \, E_{u})}\otimes Q_{y}^{({\rm b,\,E_{u}})}  \right. \\
\hfill \left. + \frac{2}{\sqrt{6}} Q_{3a}^{({\rm a,\,A_{1u}})}\otimes Q_z^{({\rm b, A_{2u}})} \right)}$  & $\braket{d,\,{\rm Zr}|p,\,{\rm O_{6i}}}$ & $\makecell{ \pm 1.67  \\ ( -1.56 \times 10^{-10}  ) }$ & $\makecell{ \mp 1.72 \times 10^{-1}  \\ ( 1.15 \times 10^{-11}  ) }$ \\ 

\hline\hline
\end{tabular*}
\renewcommand{\arraystretch}{1.0}
\end{centering}
\end{table*}

 \begin{figure*}
 \includegraphics[width=0.32\linewidth]{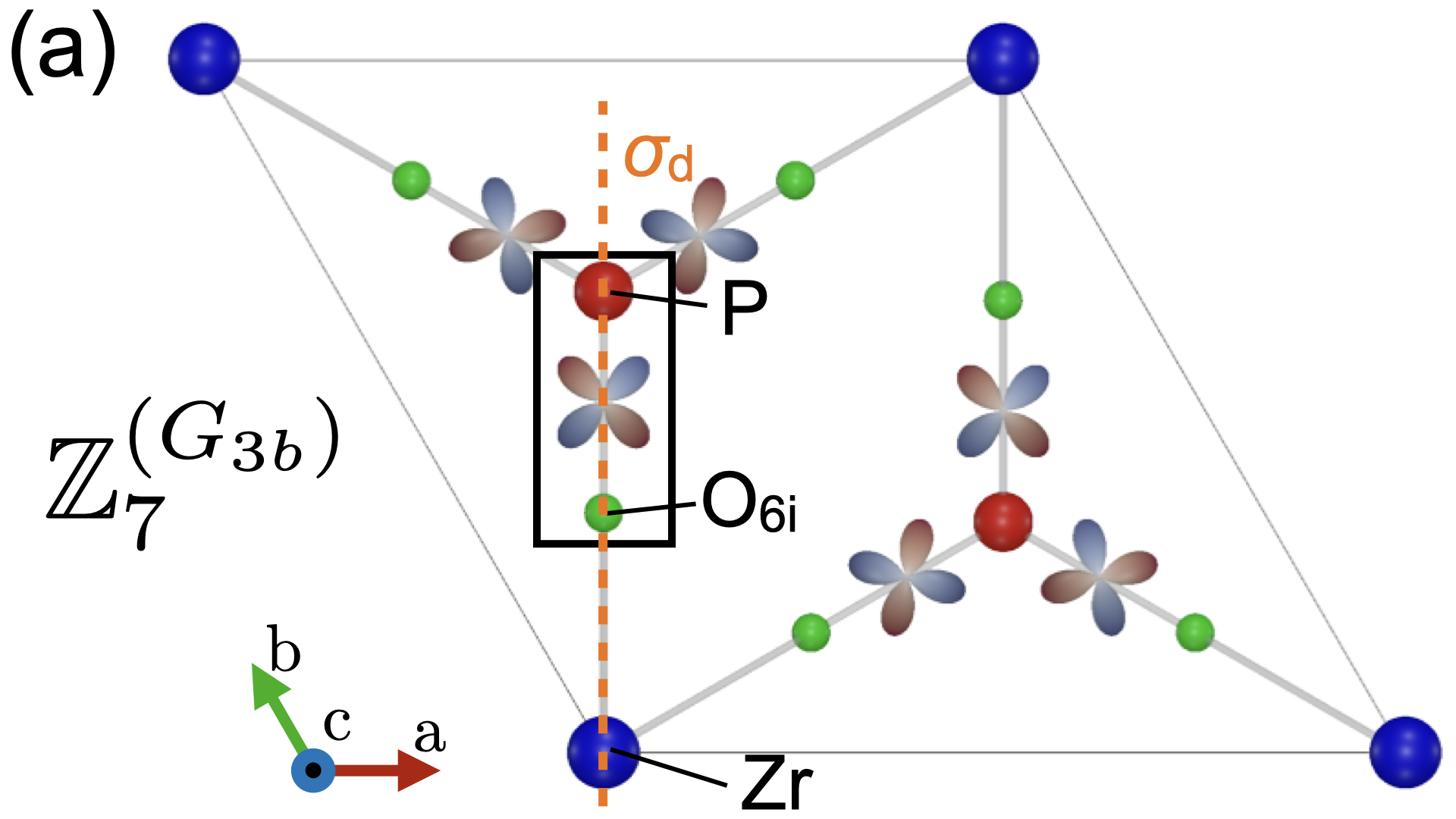} \includegraphics[width=0.32\linewidth]{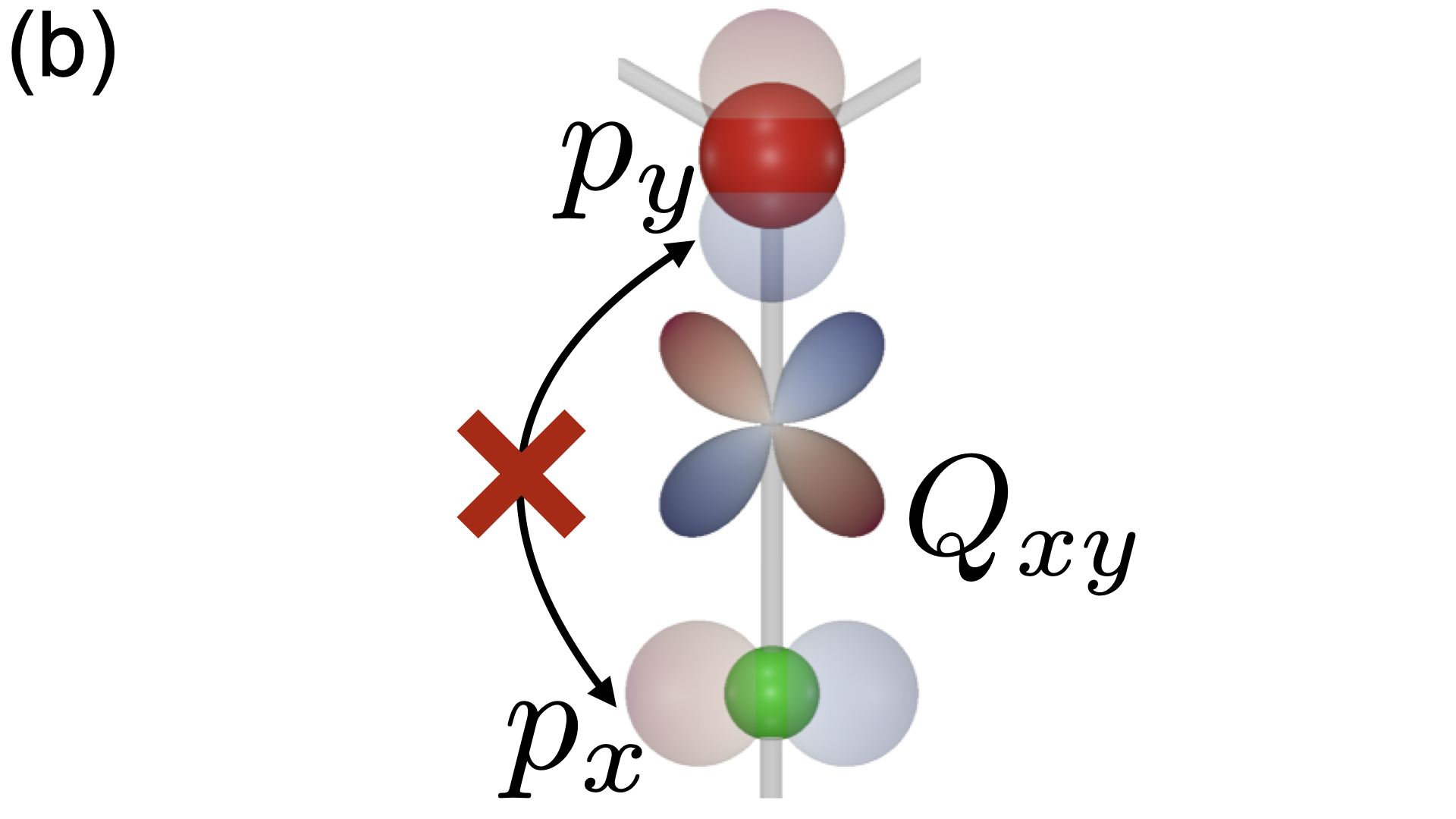} \includegraphics[width=0.32\linewidth]{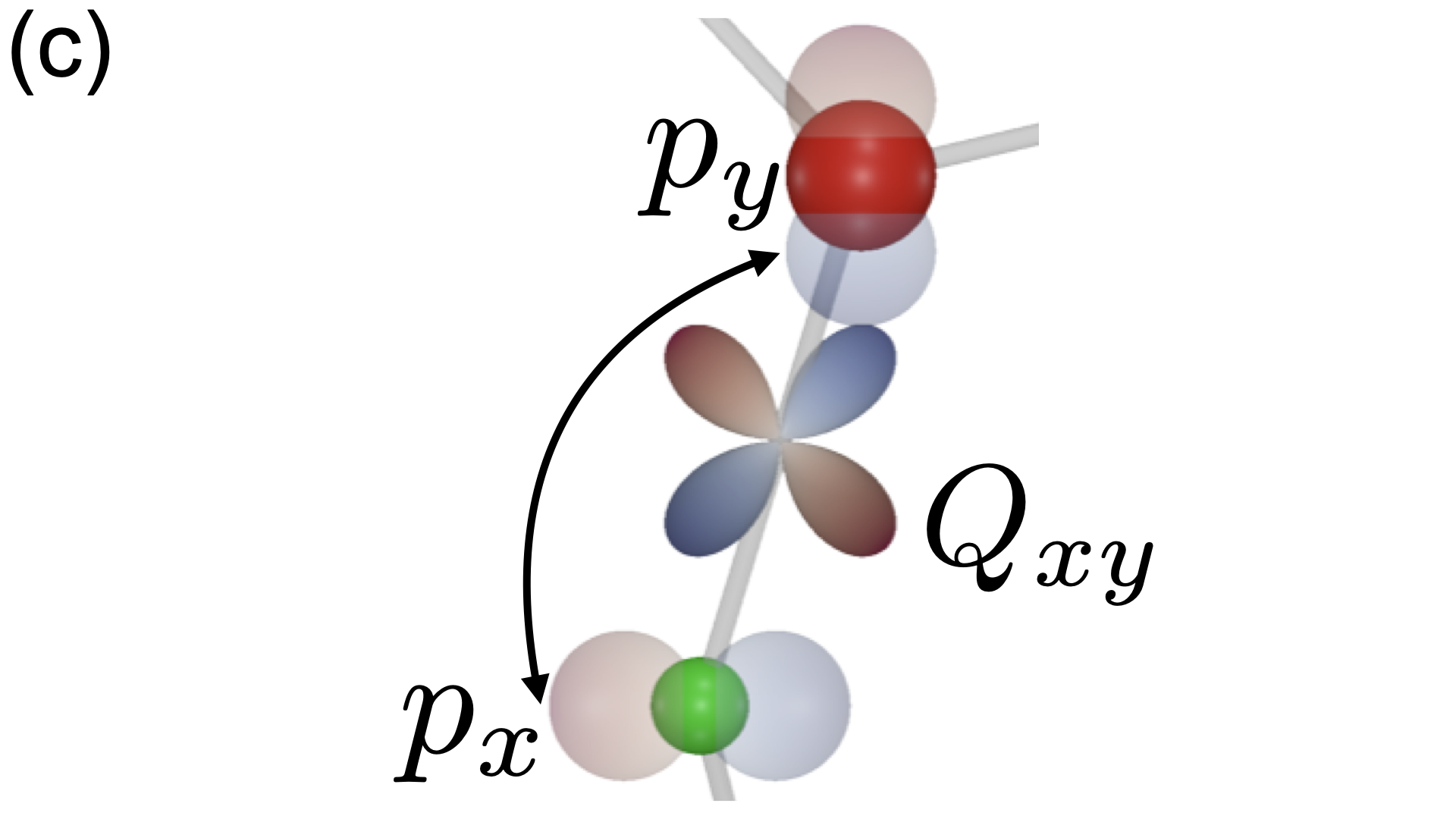}
 \includegraphics[width=0.32\linewidth]{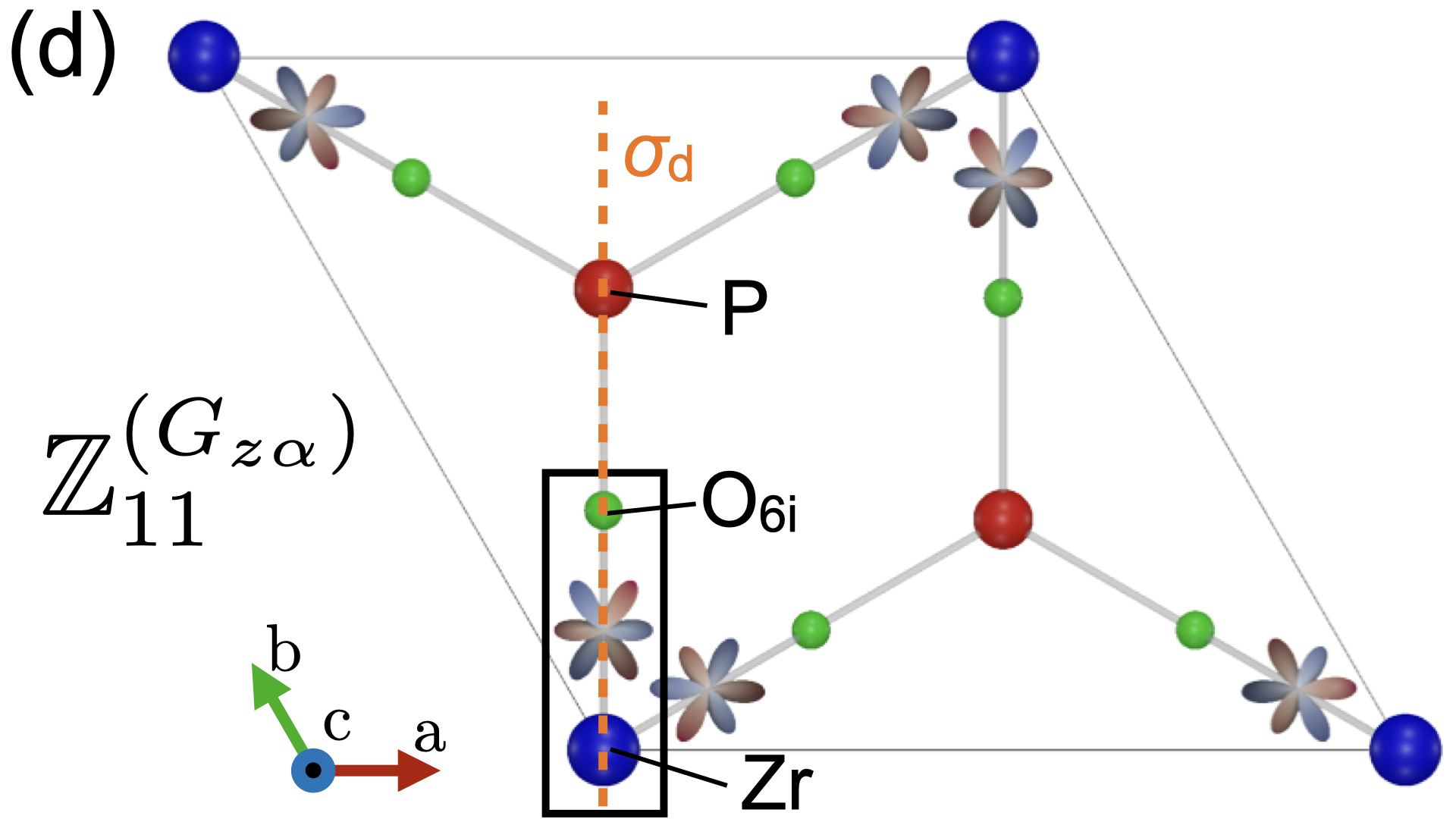} \includegraphics[width=0.32\linewidth]{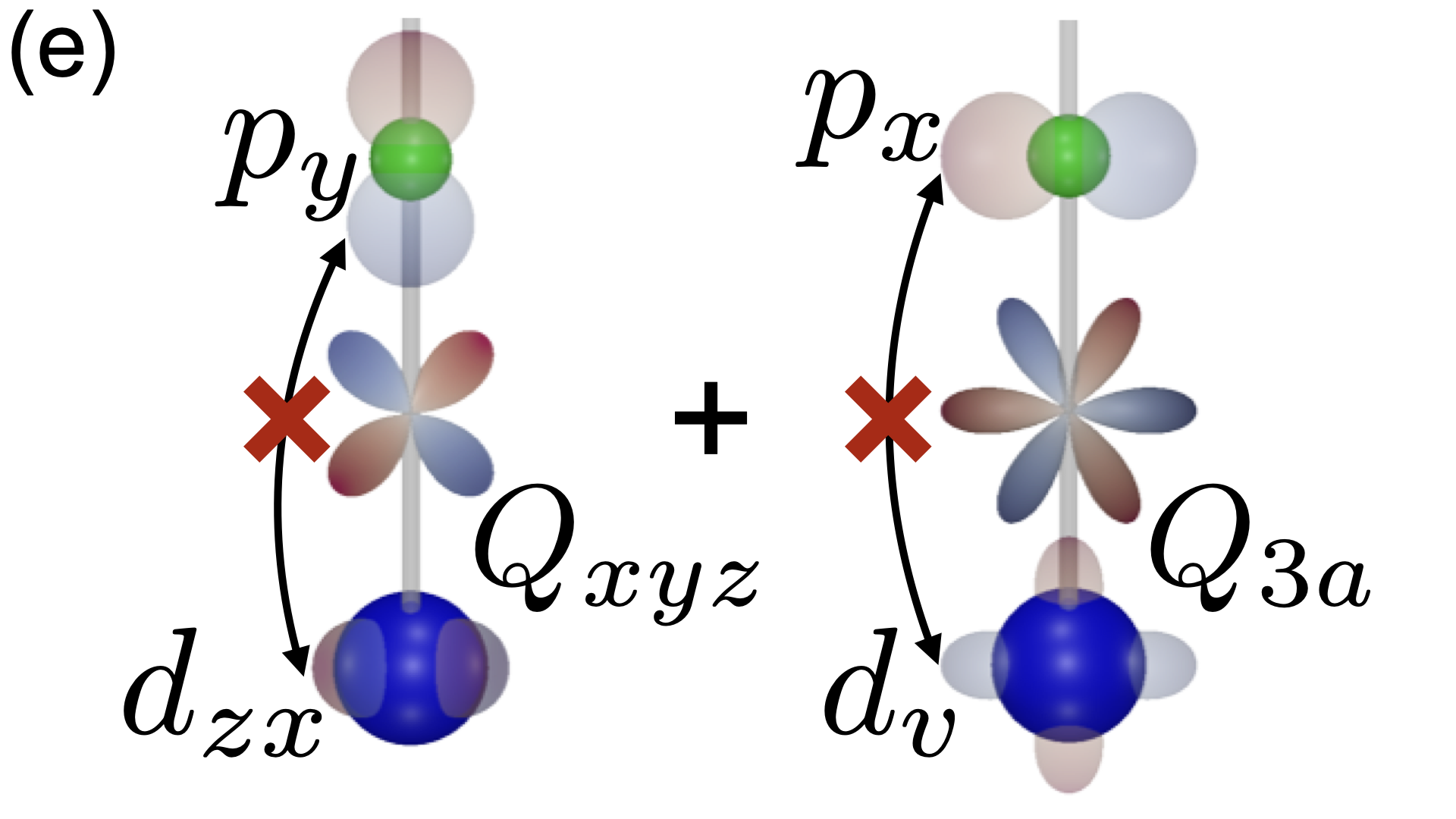} \includegraphics[width=0.32\linewidth]{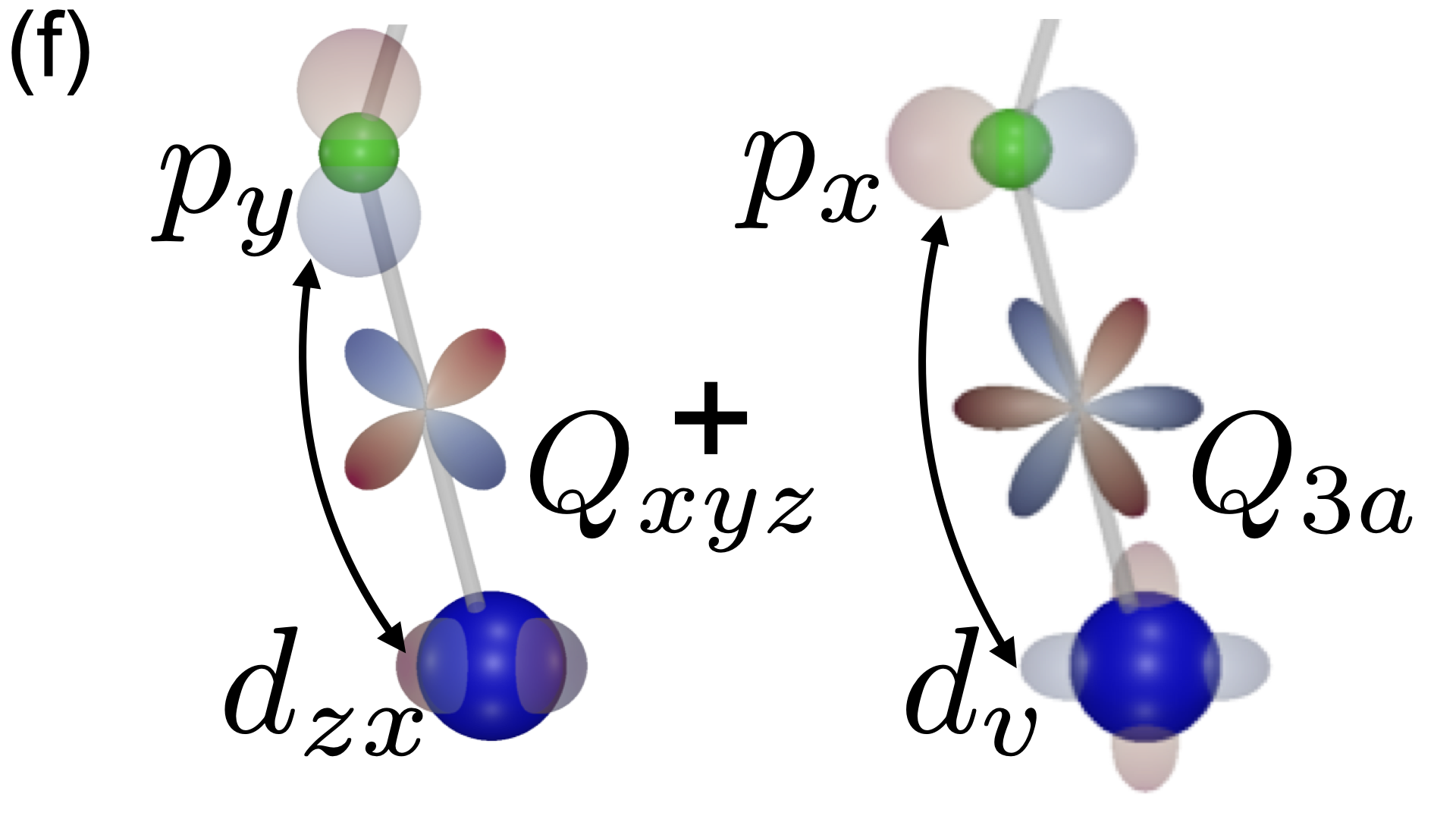}
 \includegraphics[width=0.48\linewidth]{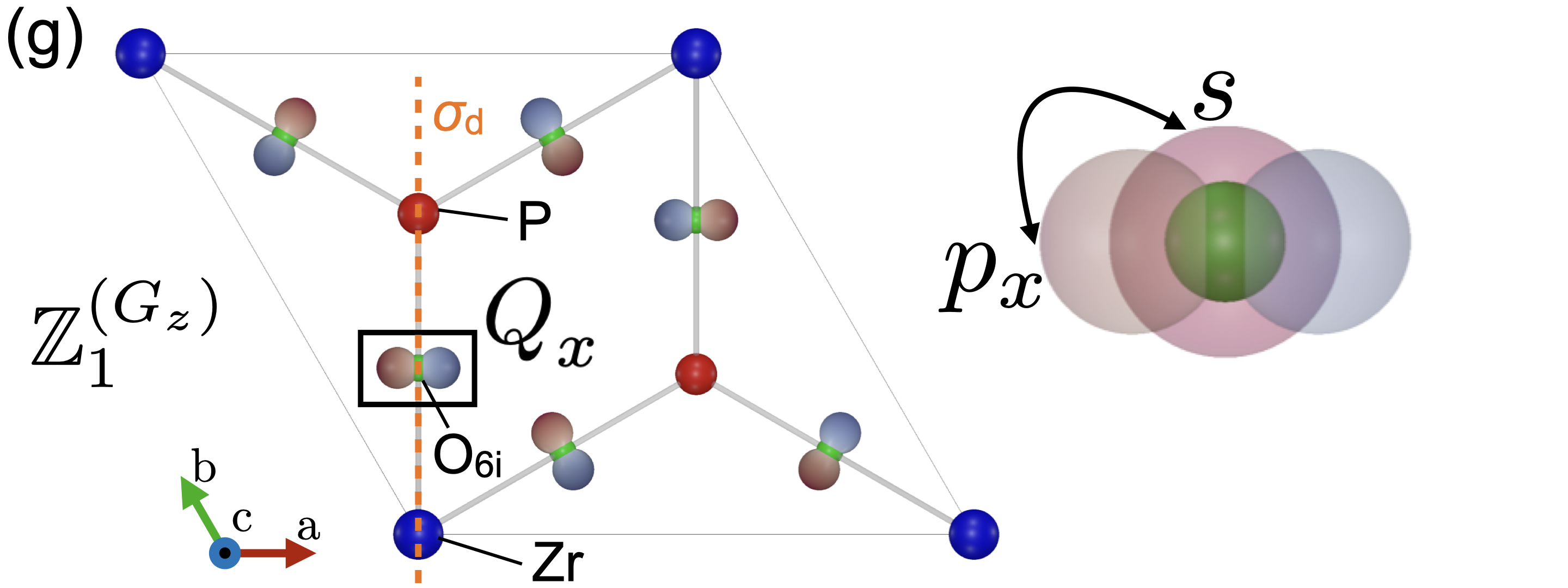} \includegraphics[width=0.48\linewidth]{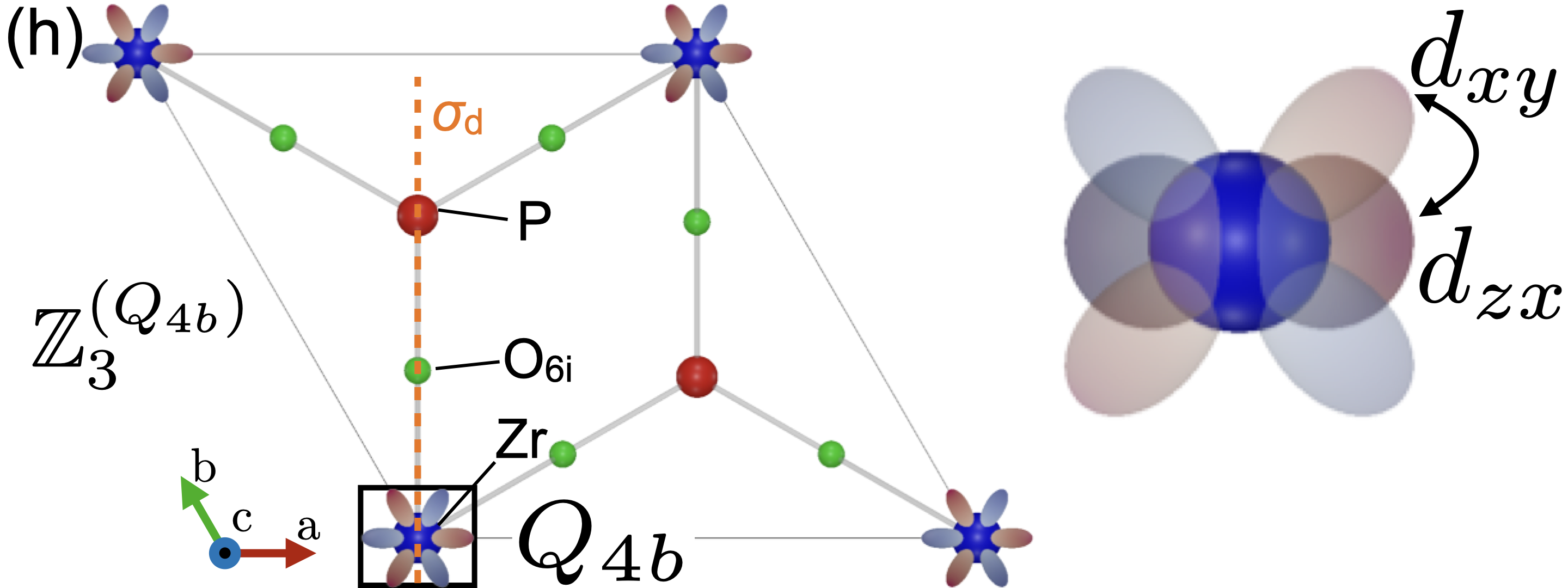}
\caption{\justifying
(a) Schematic representation of $\mathbb{Z}_{7}^{(G_{3b})}$. (b) Focusing on the region enclosed by the black rectangle in (a), $\mathbb{Z}_{7}^{(G_{3b})}$ corresponds to the spin-independent off-diagonal real hopping between the $p_{x}$ orbital at O$_{\rm 6i}$ site and the $p_{y}$ orbital at P site, (c) and this off-diagonal real hopping is allowed in the ferroaxial phase with $\varphi \neq 0$. 
In the same manner, panels (d), (e), and (f) {illustrate the} schematic representation {of $\mathbb{Z}_{11}^{(G_{z\alpha})}$}, {corresponding to} the spin-independent off-diagonal real hopping between the $p_y({\rm O_{6i}})$-$d_{zx}({\rm Zr})$ and $p_x({\rm O_{6i}})$-$d_v({\rm Zr})$ orbitals{,} {its hopping prossess} at $\varphi=0$, and {that} at $\varphi \neq 0$, respectively. 
{Panels} (g) and (h) {display the} representation and the hopping {process} of $\mathbb{Z}_{1}^{(G_{z})}$ and $\mathbb{Z}_{3}^{(Q_{4b})}${, respectively}.
}
 \label{fig_Z4}
 \end{figure*}

The explicit definitions of these SAMBs, together with their coefficients and expectation values at $\varphi = \pm 16.89^\circ,\, 0^\circ$, are summarized in Table~\ref{tab_ImportSambs}.
{In order to clarify the distinction between the multipoles that newly belong to the identity IR in $P\bar{3}$ and those that originally belong to the identity IR in $P\bar{3}m1$, we henceforth express all multipoles in terms of their irreducible representations under $P\bar{3}m1$.  
Similarly, the SAMBs are constructed with reference to the $P\bar{3}m1$ system.
}
{
{We confirm that} all the ${\rm A_{2g}}$ multipoles are negligibly small {for $\varphi=0^\circ$}, consistent with the $P\bar{3}m1$ symmetry.}

Meanwhile, {the ${\rm A_{2g}}$} multipoles take significant values {at $\varphi = \pm 16.89^\circ$} under the $P\bar{3}$ symmetry.
By comparing {the coefficients $z_{j}$ and the expectation values $\braket{{\mathbb{Z}_{j}}}$ } 
belonging to the ${\rm A_{2g}}$ IR at $\varphi = \pm 16.89^\circ$, we find that {both {$z_{j}$} and {$\braket{{\mathbb{Z}_{j}}}$} 
{of} the ETO $\mathbb{Z}_{7}^{(G_{3b})}$ in $H^{\rm (P-O_{6i}, 1)}_{\rm hop}$} {have most significant contributions}
{and those of} the ETO $\mathbb{Z}_{11}^{(G_{z\alpha})}$ in $H^{\rm (Zr-O_{6i}, 1)}_{\rm hop}$ also make noticeable contribution{s}. 
 Among the on-site CEF terms, the ETD $\mathbb{Z}_{1}^{(G_{z})}$ in $H^{\rm (O_{6i})}_{\rm CEF}$ and the EH $\mathbb{Z}_{3}^{(Q_{4b})}$ in $H^{\rm (Zr)}_{\rm CEF}$ exhibit comparatively large { coefficients  and  expectation values}.
Therefore, we focus on these four terms, $\mathbb{Z}_{7}^{(G_{3b})}$, $\mathbb{Z}_{11}^{(G_{z\alpha})}$, $\mathbb{Z}_{1}^{(G_{z})}$ and $\mathbb{Z}_{3}^{(Q_{4b})}$, and present their microscopic definitions and physical interpretations.

Figure~\ref{fig_Z4}(a) shows a schematic representation of $\mathbb{Z}_{7}^{(G_{3b})}$, corresponding to Fig.~\ref{fig_ExOfMultipoles}(e).
$\mathbb{Z}_{7}^{(G_{3b})}$ is given by the vortex structure of the spinless atomic E quadrupoles ($Q_{v}^{\rm (a)}$, $Q_{xy}^{\rm (a)}$), which are defined in the $(p_{x}, p_{y}, p_{z})$ orbitals as

\begin{align}
Q_{xy}^{({\rm a,\, E_{g}})}=\frac{\sqrt{2}}{2} \begin{pmatrix} 0&1&0\\ 1&0&0\\ 0&0&0 \end{pmatrix}, \quad
Q_{v}^{({\rm a,\, E_{g}})}=\frac{\sqrt{2}}{2} \begin{pmatrix} 1&0&0\\ 0&-1&0\\ 0&0&0 \end{pmatrix}.
\end{align}

Thus, $\mathbb{Z}_{7}^{(G_{3b})}$ represents the spin-independent off-diagonal real hopping between $p$ orbitals at nearest-neighbor P and O$_{\rm 6i}$ atoms.
As shown in Fig.~\ref{fig_Z4}(b), for the hopping between P and O$_{\rm 6i}$ atoms in the home unit-cell, $\mathbb{Z}_{7}^{(G_{3b})}$ is given by
\begin{align}
t_{1} (c_{p_{y}, {\rm P}}^{\dagger} c_{p_{x}, {\rm O_{6i}}}^{} + {\rm h.c.}),
\label{eq_Z4}
\end{align}
where $c_{p_{y}, {\rm P}}^{\dagger}$ ($c_{p_{x}, {\rm O_{6i}}}^{}$) is the creation (annihilation) operator of the electron with $p_{y}$ ($p_{x}$) orbital at P (O$_{\rm 6i}$) atom in the home unit-cell. 
Note that the hopping parameter $t_{1}$ is related to the coefficient $z_{7}$ as $t_{1} = z_{7}/\sqrt{2}$.
As shown in Figs.~\ref{fig_Z4}(b) and (c), this hopping is allowed only in the ferroaxial phase with $\varphi \neq 0$ by breaking the mirror symmetry $\sigma_{\rm d}$ shown in Fig.~\ref{fig_structure}.
Indeed, as shown in Table~\ref{tab_ImportSambs}, $z_{7}$ is close to zero when $\varphi = 0^\circ$, while $z_{7} {= \pm 7.21~{\rm eV}}$ in the ferroaxial phase {with $\varphi = \pm 16.89^\circ$}, and its sign corresponds to the direction of the rotational distortion.

Additionally, the expectation value $\braket{\mathbb{Z}_{7}^{(G_{3b})}}$ at $\varphi = \pm 16.89^\circ$ has most significant contribution, $\braket{\mathbb{Z}_{7}^{(G_{3b})}} = \mp 4.60 \times 10^{-1}$.

Similarly, $H^{\rm (Zr-O_{6i}, 1)}_{\rm hop}$ contains another ETO $\mathbb{Z}_{11}^{(G_{z\alpha})}$, corresponding to Fig.~\ref{fig_ExOfMultipoles}(b).
Figure~\ref{fig_Z4}(d) shows a schematic representation of $\mathbb{Z}_{11}^{(G_{z\alpha})}$, and the spinless atomic E octupoles ($Q_{z\beta}^{\rm (a)}$, $Q_{xyz}^{\rm (a)}$, $Q_{3a}^{\rm (a)}$) that constitute $\mathbb{Z}_{11}^{(G_{z\alpha})}$ are defined in the $p$--$d$ hybrid orbitals space $\braket{p|d}$ as
\begin{align}
&Q_{z\beta}^{({\rm a,\, E_u})}=\frac{\sqrt{3}}{3}
\begin{pmatrix}
0&0&0&-1&0\\
0&0&1&0&0\\
0&-1&0&0&0
\end{pmatrix}
,\\
&Q_{xyz}^{({\rm a,\, E_u})}=\frac{\sqrt{3}}{3}
\begin{pmatrix}
0&0&1&0&0\\
0&0&0&1&0\\
0&0&0&0&1
\end{pmatrix}
, \\
&Q_{3a}^{({\rm a,\, A_{1u}})}=\frac{\sqrt{2}}{2}
\begin{pmatrix}
0&1&0&0&0\\
0&0&0&0&-1\\
0&0&0&0&0
\end{pmatrix}
, 
\end{align}
where the rows and columns correspond to the $(p_x, p_y, p_z)$ orbitals and the $(d_{{u}}, d_{{v}}, d_{yz}, d_{zx}, d_{xy})$ orbitals {with $v=x^2-y^2$}, respectively.
{Accordingly,} $\mathbb{Z}_{11}^{(G_{z\alpha})}$ represents the spin-independent off-diagonal real hopping between $d$ orbitals on Zr atoms and $p$ orbitals on O$_{\rm 6i}$ atoms.
As shown in Fig.~\ref{fig_Z4}(e), for the hopping between Zr and O$_{\rm 6i}$ atoms in the home unit-cell, $\mathbb{Z}_{11}^{(G_{z\alpha})}$ is given by
\begin{align}
t_{2} (c_{d_{zx}, {\rm Zr}}^{\dagger} c_{p_{y}, {\rm O_{6i}}}^{} + {\rm h.c.}) + t_{3} (c_{d_{v}, {\rm Zr}}^{\dagger} c_{p_{x}, {\rm O_{6i}}}^{} + {\rm h.c.}),
\label{eq_Z4}
\end{align}
where $c_{d_{v}, {\rm Zr}}^{\dagger}$ ($c_{p_{y}, {\rm O_{6i}}}^{}$) is the creation (annihilation) operator of the electron with $d_{v}$ ($p_{y}$) orbital at Zr (O$_{\rm 6i}$) atom in the home unit-cell. 
The hopping parameters $t_{2}$ and $t_{3}$ are related to the coefficient $z_{11}$ as $t_{2} = z_{11}/6\sqrt{2}$ and $t_{3} =  z_{11}/4$, respectively.
Similarly to $\mathbb{Z}_{7}^{(G_{3b})}$, this hopping is allowed only in the ferroaxial phase with $\varphi \neq {0^\circ}$ by breaking the mirror symmetry $\sigma_{\rm d}$, as shown in Figs.~\ref{fig_Z4}(e) and (f).
This is confirmed from the DFT calculations, where $z_{11}$ is {close to} zero when $\varphi = 0^\circ$, while $z_{11} = \pm {4.35}$ {eV} in the ferroaxial phase {with nonzero $\varphi$}, and its sign is reversed according to the direction of the rotational distortion{, as shown in Table~\ref{tab_ImportSambs}}.
The expectation value $\mathbb{Z}_{11}^{(G_{z\alpha})}$ at $\varphi = \pm 16.89^\circ$ is the second largest following $\braket{\mathbb{Z}_{7}^{(G_{3b})}}$, $\braket{\mathbb{Z}_{11}^{(G_{z\alpha})}} = \mp 3.80 \times 10^{-1}$.

{I}n the CEF terms, the ETD $\mathbb{Z}_{1}^{(G_{z})}$ in $H^{\rm (O_{6i})}_{\rm CEF}$ and the EH $\mathbb{Z}_{3}^{(Q_{4b})}$ in $H^{\rm (Zr)}_{\rm CEF}$ also make noticeable contributions.
As shown in Fig.~\ref{fig_Z4}(g),  the site-cluster ETD $\mathbb{Z}_{1}^{(G_{z})}$ arises from the {vortex-like configuration of the} spinless atomic E dipoles $\bm{Q}^{\rm (a)} {= (Q^{\rm (a)}_{x}, Q^{\rm (a)}_{y})}$ on the O$_{\rm 6i}$ atoms, 
{which are defined in $\braket{s|p}$ orbitals space as 
\begin{align}
Q_{x}^{({\rm a,\, E_{u}})}= \begin{pmatrix} 1&0&0 \end{pmatrix}, \quad
Q_{y}^{({\rm a,\, E_{u}})}= \begin{pmatrix} 0&1&0 \end{pmatrix}{.}
\end{align}
}
This ETD is {also} allowed only in the ferroaxial phase {with $\varphi \neq 0$}, leading to {the on-site} hybridization between the $s$ and $(p_x, p_y)$ orbitals at the O$_{\rm 6i}$ sites{.}
On the other hand, $\mathbb{Z}_{3}^{(Q_{4b})}$ corresponds to the atomic EH $Q_{4b}^{\rm (a)}$ on the Zr atom, 
{
which are defined in the $(d_u,\,d_v,\,d_{xy},\,d_{yz},\,d_{zx})$ orbitals:
\begin{align}
&Q_{4b}^{({\rm a,\, A_{2g}})}=\frac{1}{2}
\begin{pmatrix}
0&0&0&0&0\\
0&0&0&-1&0\\
0&0&0&0&1\\
0&-1&0&0&0\\
0&0&1&0&0
\end{pmatrix}
.
\end{align}
}
As shown in Fig.~\ref{fig_Z4}(h), {$\mathbb{Z}_{3}^{(Q_{4b})}$ corresponds to} the {$d_{v}$-$d_{yz}$ and $d_{zx}$-$d_{xy}$ hybridization between the $d$} orbitals at the Zr atom. The signs of the coefficients $z_{1}$ and $z_{3}$, as well as those of the expectation values $\braket{\mathbb{Z}_{1}^{(G_{z})}}$ and $\braket{\mathbb{Z}_{3}^{(Q_{4b})}}$, correspond to the direction of the rotational distortion.

From the above results, we conclude that the {two} ETOs $\mathbb{Z}_{7}^{(G_{3b})}$, corresponding to the spin-independent off-diagonal real hopping between $p$ orbitals at nearest-neighbor P and O$_{\rm 6i}$ atoms, and $\mathbb{Z}_{11}^{(G_{z\alpha})}$, corresponding to the spin-independent off-diagonal real hopping between $d$ orbitals on Zr atoms and $p$ orbitals on O$_{\rm 6i}$ atoms, provide the dominant contributions to the ferroaxial transition.
In addition, the ETD $\mathbb{Z}_{1}^{(G_{z})}$ in $H^{\rm (O_{6i})}_{\rm CEF}$ and the EH $\mathbb{Z}_{3}^{(Q_{4b})}$ in $H^{\rm (Zr)}_{\rm CEF}$, have also significant contribution to the CEF terms in the ferroaxial phase.

\begin{figure*}
 \includegraphics[width=1\linewidth]{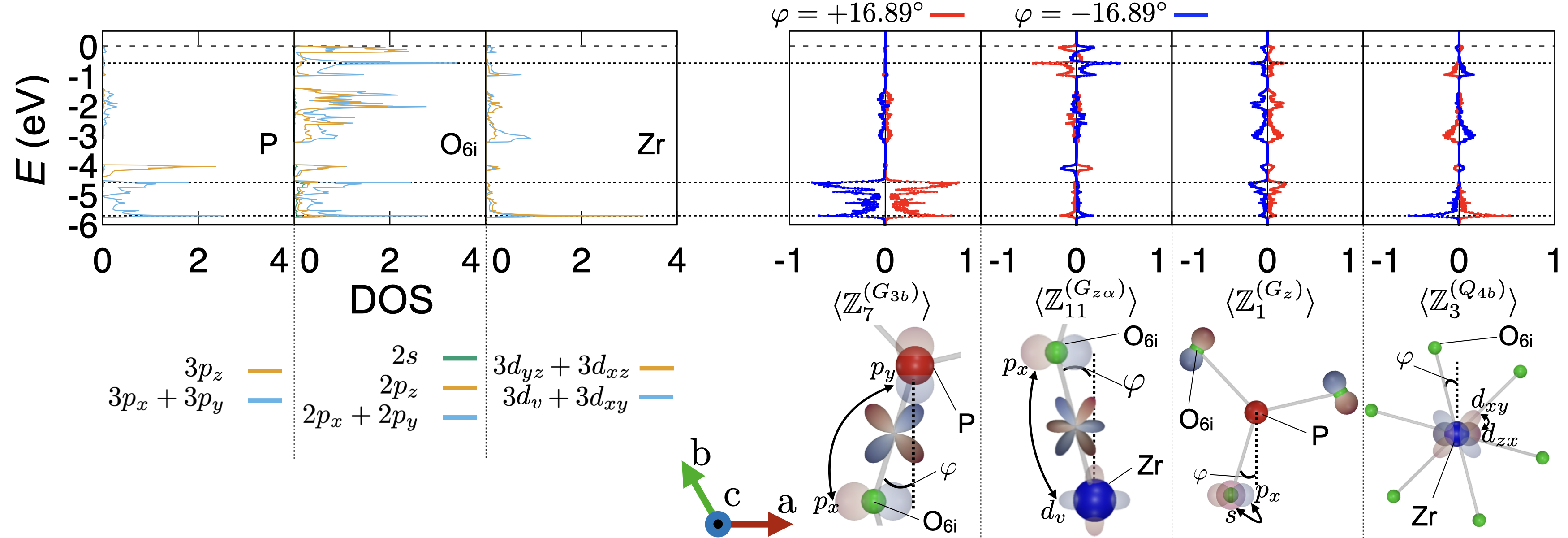}
\caption{\justifying
Energy dependence of the
expectation values $\braket{\mathbb{Z}_j}$ {(right panel)} together with the DOS {(left panel)} obtained from the DFT calculations {in} Fig.~\ref{fig_bandDOS}(b). 
The red and blue lines represent the results for $\varphi = +16.89^\circ$ and $\varphi = -16.89^\circ$, respectively. 
}
 \label{fig_dosDetail}
\end{figure*}

To further clarify the orbital characters involved in the relevant hybridizations associated with the above four multipoles, we evaluate the {energy-resolved} expectation values $\braket{\mathbb{Z}_j}$ for $\varphi = \pm16.89^\circ$.
Figure~\ref{fig_dosDetail} shows the energy dependence of $\braket{\mathbb{Z}_{7}^{(G_{3b})}}$,  $\braket{\mathbb{Z}_{11}^{(G_{z\alpha})}}$, $\braket{\mathbb{Z}_{1}^{(G_{z})}}$, and $\braket{\mathbb{Z}_{3}^{(Q_{4b})}}$ together with the DOS {in} Fig.~\ref{fig_bandDOS}({a}). They take nonzero values in the ferroaxial phase, and their signs reverse according to the direction of the rotational distortion, indicating that these multipoles serve as quantitative measures of the ferroaxiality.
Notably, the signs of the bond-cluster ETO $\braket{\mathbb{Z}_{7}^{(G_{3b})}}$  and the site-cluster ETD $\braket{\mathbb{Z}_{1}^{(G_{z})}}$  are almost uniquely determined, whereas the signs of the bond-cluster ETO $\braket{\mathbb{Z}_{11}^{(G_{z\alpha})}}$ and the atomic EH $\braket{\mathbb{Z}_{3}^{(Q_{4b})}}$ reverse depending on the energy level.

Furthermore, each peak appearing in $\braket{\mathbb{Z}_j}$ corresponds to a peak in the DOS of the relevant orbitals involved in the hybridization associated with $\mathbb{Z}_j$. 
Once the ferroaxial order appears, {additional} off-diagonal hybridizations among the $p$--$p$, $s$--$p$, and $p$--$d$ orbitals emerge.
These hybridizations induce new bonding and antibonding states, giving rise to additional peaks in the DOS that are absent in the nonferroaxial phase.

\begin{figure*}
  \includegraphics[width=0.9\linewidth]{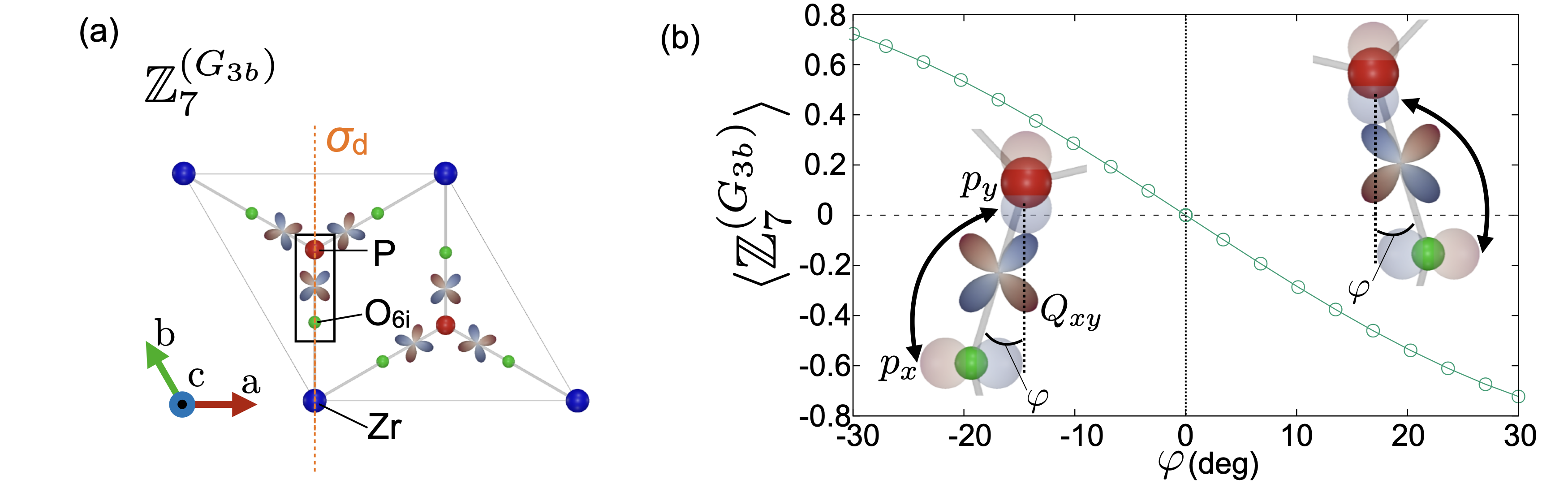}
 \includegraphics[width=0.9\linewidth]{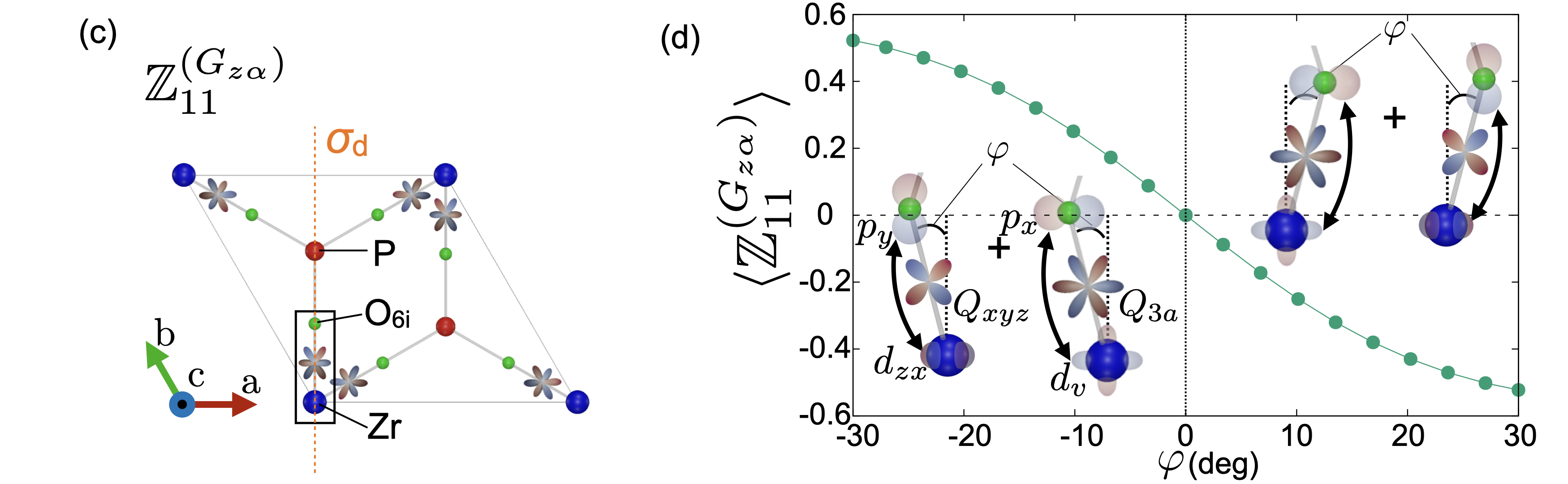}
   \includegraphics[width=0.9\linewidth]{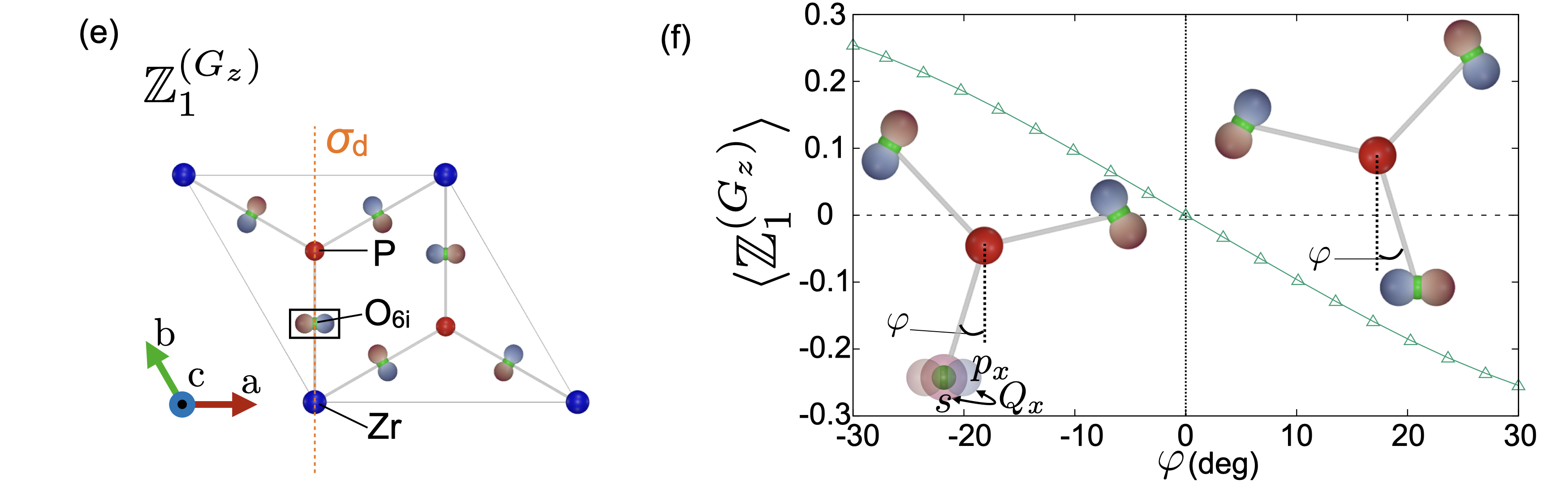} 
   \includegraphics[width=0.9\linewidth]{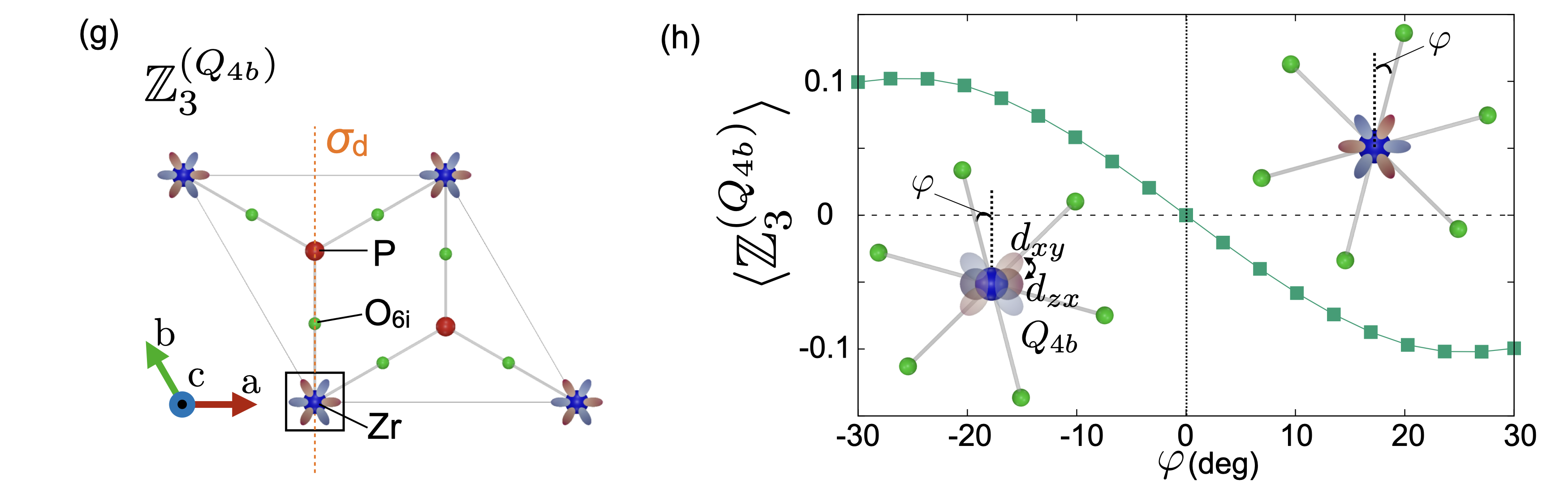}
\caption{\justifying
SAMBs belonging to the ${\rm A_{2g}}$ IR are shown in the left column: those located at (a) the ${\rm P\!-\!O_{6i}}$ bond (${\mathbb{Z}_{7}^{(G_{3b})}}$), (c) the ${\rm Zr\!-\!O_{6i}}$ bond (${\mathbb{Z}_{11}^{(G_{z\alpha})}}$), (e) the ${\rm O_{6i}}$ site (${\mathbb{Z}_{1}^{(G_{z})}}$), and (g) the ${\rm Zr}$ site (${\mathbb{Z}_{3}^{(Q_{4b})}}$).  The mirror symmetry $\sigma_{\rm d}$, indicated by the orange dashed line, forbids these hybridizations at $\varphi = 0^\circ$. 
{W}hen the symmetry is reduced to ${P\bar{3}}$, $\sigma_{\rm d}$ is broken, and the SAMBs become {nonzero}, as shown in the right column. The variations in the expectation values of the SAMBs respect to $\varphi$ are also shown in the right column.}
 \label{fig_Zs}
 \end{figure*}

\subsection{{Rotation angle dependence of} ferroaxiality}

We show the evolution of the expectation values of the four SAMBs by systematically changing the rotation angle $\varphi$ from {$0^{\circ}$} to $\pm 30^\circ$.
Figure~\ref{fig_Zs} presents how $\braket{\mathbb{Z}_j}$ varies with the rotational angle $\varphi$. 
As shown in Figs.~\ref{fig_Zs}(b), (d), (f), and (h), $\braket{\mathbb{Z}_j}$ vanishes in the nonferroaxial phase with the mirror symmetry $\sigma_{\rm d}$ {($\varphi={0^{\circ}}$)}, whereas it becomes finite in the ferroaxial phase {($\varphi \neq {0^{\circ}}$)} and increases linearly with increase of $\varphi$ from zero.
Furthermore, The sign of $\braket{\mathbb{Z}_j}$ reverses depending on the direction of rotation.
Therefore, $\braket{\mathbb{Z}_{7}^{(G_{3b})}}$, $\braket{\mathbb{Z}_{11}^{(G_{z\alpha})}}$, $\braket{\mathbb{Z}_{1}^{(G_{z})}}$, and $\braket{\mathbb{Z}_{3}^{(Q_{4b})}}$ play as proper indicators of ferroaxiality, and their sign correspond to the direction of rotation.

\subsection{Effect of relativistic spin–orbit coupling}
\label{sec: Effect of relativistic spin–orbit coupling}
Finally, we discuss the effect of the relativistic spin-orbit coupling on the ferroaxial transition, especially the contribution of the spinful ETD represented by the cross-product--type {SOC} {${G_z'}= (\bm{l} \times \bm{\sigma})_z / \sqrt{2}$}~\cite{Hayami_PhysRevB.98.165110}, where $\bm{l}$ and $\bm{\sigma}$ denote the {atomic} orbital and spin angular momenta, respectively.
We also performed DFT calculation by using the fully-relativistic ONCV pseudopotentials~\cite{ONCV_PRB_2013} downloaded from PseudoDojo \cite{VANSETTEN201839} at $\varphi=\pm 16.89^\circ$, and built spinful SymCW models. 
All the parameters used in the DFT calculation keeps as same as the spinless calculation with the scalar-relativistic ONCV pseudopotentials.
{
By considering the SOC, the additional SOC{-}driven local term $H_{\rm SOC}$ is appeared in the SymCW Hamiltonian given by Eq.~(\ref{eq_H_SymCW_2}), and it is expressed as the linear combination of the product of the spinful SAMB $\mathbb{Z}_{j}'$ and its coefficient $z_{j}'$, where the prime denotes the spinful term. 
}
Using the obtained spinful SymCW models, we evaluated {the coefficient ${z_j'}$ and the expectation value $\braket{\mathbb{Z}_j'}$} of the {spinful}
ETD ${G_z'}$
at {P, Zr, O$_{\rm 6i}$ atoms}, which are summarized in Table.~\ref{tab_spinfulres}.

As shown in Table.~\ref{tab_spinfulres}, the contributions of the spinful ETDs are orders of magnitude smaller than those of the spinless ferroaxial terms {locate{d} at {the} same {atom} } in Table.~\ref{tab_ImportSambs}. 
We further examined the other spinful contributions including spin-dependent hoppings, and confirmed that their magnitudes are below {$4\times10^{-2}$} and  $2\times10^{-3}$ {for ${z_j'}$ and ${\braket{\mathbb{Z}_j'}}$, respectively}. 
{In addition, we confrimed that spinless ferroaxial SAMBs, such as $z_7$ and $z_{11}$, are quantitatively almost unchanged.}
Therefore, we conclude that the relativistic {SOC} has a negligible influence on the ferroaxial transition.

\begin{table}
{
\setlength{\tabcolsep}{12pt}
\caption{\justifying
{Coefficient $z_j'$ and expectation value $\braket{\mathbb{Z}_j'}$  of the site-cluster ETD at }
P, Zr, and O$_{\rm 6i}$ atoms.
The column of ``Hybridization" represents the corresponding Hilbert space.
}
\label{tab_spinfulres}
\renewcommand{\arraystretch}{1.5}
\begin{tabular}{ccc}
\hline\hline
Hybridization & {$z_j^{(G_z^{\rm spinful})}~[{\rm eV}]$} & {$\braket{\mathbb{Z}_j^{(G_z^{\rm spinful})}}$} \\
 \hline
{$\braket{p, {\rm P}|p, {\rm P}}$} & {$9.33\times10^{-5}$} & $1.88\times10^{-4}$ 
\\
{$\braket{d, {\rm Zr}|d, {\rm Zr}}$} & {$-1.05\times10^{-3}$} & $1.65\times10^{-6}$ 
\\
{$\braket{p, {\rm O_{6i}}|p, {\rm O_{6i}}}$} & {$-1.14\times10^{-2}$} & $1.06\times10^{-3}$ 
\\
\hline\hline
\end{tabular}
\setlength{\tabcolsep}{6pt}
}
\end{table}

\section{CONCLUSION}
\label{sec_conc}
We have performed a comprehensive microscopic analysis of the displacive-type ferroaxial transition in K$_2$Zr(PO$_4$)$_2$ by combining DFT calculations with a SymCW approach formulated in the framework of SAMB. 
This {method} has enabled us to quantify the electronic ferroaxial degrees of freedom directly from realistic band structures and to decompose the Hamiltonian into a {linear combination of the} complete set of spinless multipoles.
Within this multipole-based description, we have identified that the ETD, the ETO, and the EH, all sharing the same symmetry, emerge as the key microscopic ingredients of the ferroaxial phase. 
Among them, the bond-cluster ETOs associated with the P-O$_{\rm 6i}$ and Zr-O$_{\rm 6i}$ bonds provide the dominant contributions: they arise from spin-independent off-diagonal real hoppings between the $p$ orbitals on P and O atoms and between the $d$ orbitals on Zr atoms and the $p$ orbitals on O atoms. 
We have further shown that the site-cluster ETD at the O$_{\rm 6i}$ sites and the atomic EH at the Zr sites also make non-negligible contributions through on-site hybridizations between $s$ and $(p_x, p_y)$ orbitals and between different $d$ orbitals, respectively. 
The expectation values of these multipoles vanish in the nonferroaxial $P\bar{3}m1$ phase, while those become finite in the ferroaxial $P\bar{3}$ phase, and grow approximately linearly with the rotation angle $\varphi$, with their signs following the direction of the rotational distortion. 
They therefore serve as quantitative order parameters for the electronic aspect of ferroaxiality.
Furthermore, we have demonstrated that the effect of relativistic {SOC} is negligibly small in the ferroaxial transition.

Overall, our results demonstrate that the essential electronic-origin ferroaxiality in K$_2$Zr(PO$_4$)$_2$ arises predominantly from spin-independent orbital hybridizations among different orbitals and different atoms, rather than from relativistic {SOC}. 
The present analysis highlights the power of the SAMB-based SymCW framework for disentangling microscopic ferroaxial degrees of freedom in real materials, and it provides a useful basis for exploring ferroaxiality and related rotational structural transitions in a broader class of compounds.

{
\begin{acknowledgments}
The authors thank H. Kusunose for fruitful discussions.
This work {was supported by JSPS KAKENHI Grants Numbers JP22H00101, JP22H01183, JP23H04869, JP23K03288, and by JST CREST (JPMJCR23O4) and JST FOREST (JPMJFR2366).
} 
 \end{acknowledgments}
 }

\bibliography{refs}

\end{document}